\documentclass[10pt,authoryear]{elsarticle}
\pdfoutput=1
\usepackage{graphicx}
\usepackage{amsmath}
\usepackage{amssymb}
\usepackage{url,a4wide,fullpage}
\newcommand{\pd}[3]{\frac{\partial^{#3} #1}{\partial {#2}^{#3}}} 
\newcommand{\td}[3]{\frac{d^{#3} #1}{d {#2}^{#3}}} 
\renewcommand{\v}[1]{\ensuremath{\mathbf{#1}}} 
\newcommand{\gv}[1]{\ensuremath{\mbox{\boldmath$ #1 $}}}
\renewcommand{\bar}[1]{\ensuremath{\overline{#1}}}
\newcommand{\msol}{M$_{\odot}\,$}
\newcommand{\sigv}{\ensuremath{\langle \sigma V \rangle}}

\journal{Astroparticle Physics}

\begin{document}
	
\begin{frontmatter}
\title{Hunting Dark Matter in ultra-compact structures within the Milky-Way.}
\author[a]{G.~Beck\corref{corr1}}
\ead{geoffrey.beck@wits.ac.za}
\author[a]{S.~Colafrancesco}
\ead{sergio.colafrancesco@wits.ac.za}
\address[a]{School of Physics, University of the Witwatersrand, Private Bag 3, WITS-2050, Johannesburg, South Africa}

\cortext[corr1]{Corresponding author}

\begin{abstract}
	The local environment is ideal for searching out compact dark structures via the microlensing and multi-frequency emissions as these objects are expected to be faint and microlensing experiments have already hinted at their possibility. In the case that these objects are composed of Dark Matter (DM) then there are both few limits on their abundance but many consequences of their existence or non-existence on both local and cosmic scales.
	
	In this work we examine the possibility of Ultra-Compact Mini-Halos (UCMHs) formed in the early universe. These objects can persist to the present epoch due to their large central density inuring them to the worst effects of later tidal stripping. As such, these objects could constitute probes of many details of early universe physics, such as primordial phase-transitions, inflation, small scale exploration of the primordial density perturbation field and non-Gaussianity of these density perturbations. The fact that they are also highly dark matter-dominated objects means that they are attractive objects of study in the continuing hunt for the nature of Dark Matter (DM) through indirect detection. Another reason to study such objects in the local environment is found in the conjecture that encounters with UCMHs could induce catastrophic events on planets within our solar system, e.g. mass-extinction events on Earth.
	
	We will outline a strategy for multi-frequency UCMH searches within the region of the Milky-Way in which Gaia can accurately reconstruct microlens masses. This methodology ensures that any candidate UCMH DM emission should correlate to some unidentified microlensing object with determinable mass.
	By means of the projected sensitivities of the upcoming SKA and CTA experiments we then determine the future prospects for constraining the parameter space of UCMH objects and show that these projects can greatly extend the current scope of the Fermi-LAT constraints.
	
	We further studied the impact of these systems on the extreme scales in the universe: cosmological scales and local planetary environments within the solar system.
	We found that, using our abundance limits, UCMHs can set strong constraints on the running of the power spectrum of primordial fluctuation spectrum of, $\beta_s \lesssim -0.0898 \alpha_s + 0.003622$.
	On solar system scales, we revised the estimates of the DM-induced heating of the Earth's core by encounters with UCMHs, and we show that the results obtained in the literature are unrealistic, thus casting strong doubts on whether DM can induce mass extinction events during the past history of Earth.
	
	However, we find that the DM-induced heating can have a more substantial effect in the core of Mars, with the possibility to then de-gas its mantle and produce a quenching of the geodynamo. This DM-induced effect could provide an explanation for the loss of the Mars atmosphere and for the absence of water on the planet. This hypothesis is found to be unviable for 1 TeV WIMPs due to Fermi-LAT UCMH abundance constraints, but lower masses annihilating via quarks are shown to remain far less constrained, even by the up-coming SKA.
\end{abstract}


\end{frontmatter}


\section{Introduction}
The local environment of the Milky-Way around our solar system is one suggested to contain a multitude of unknown planets and other compact objects. This is indicated by the hints that emerge from microlensing optical depth measurements by EROS-2~\citep{tisserand2006} as well as the MACHO collaboration~\citep{alcock2000b} and later revisions of their data~\citep{bennett2005}. Numerous candidate microlensing events by possible compact structures within the Milky-Way have also been listed by a multitude of experiments~\citep{alcock2000b,lenses1,lenses2,lenses3,lenses4,lenses5,tisserand2006,lensing1,lensing2}. More recently those of the MOA and OGLE experiments~\citep{moa-data1,lensing3} and with early hints in the first Gaia~\citep{gaia-docs} data releases~\citep{gaia-data1}. These were sufficient to rule out compact structures between $10^{-7}$ and $15$ M$_{\odot}$ as the majority of Dark Matter (DM) but allow for as much as $\mathcal{O}$(10\%) of the Milky-Way DM mass to be so constituted. Gaia is expected to vastly expand the hunt for local microlensing objects~\citep{gaia2001,herzog2017} due its capacity to perform precision astrometry on millions of local stars, putting it on the fore-front of future searches for these objects, this has been demonstrated particularly for ultra-compact dark objects on small scales~\citep{li2012}. Thus, it's capabilities will be used in this work to benchmark the region in which we can hope to accurately determine the masses of these objects. Importantly, Gaia has already detected substructure within the Milky-Way halo~\citep{myeong2017}, and been used to characterise DM velocities~\citep{herzog2017}, with the expectation of more data allowing far finer searches for DM (and other compact) substructure~\citep{feldmann2015,herzog2017}. Additional emphasis can be placed upon the hunt for compact micro-structures by recent results \citep{new-lens}, using multi-frequency data to discover long-term variation of light curves in active galactic nuclei. This data has hinted at the presence of microlensing objects between $10^3$ and $10^6$ M$_{\odot}$. Given the relative weakness of constraints on ultra-compact DM structures, the possibility of unknown microlensing objects should motivate a thorough investigation of their possible consequences.

There are several candidates to explain a large enough population of compact dark/faint objects in the Milky-Way to account for the candidate lensing events. Examples being brown-dwarf~\citep{evans2014} and ultra-cool stars (both already being probed by Gaia~\citep{smart2017}) and primordial black holes, which have a history of constraint via microlensing~\citep{carr2016}.
However, one intriguing candidate for such objects are Ultra-Compact Mini-Halos (UCMHs). These are conjectured to be composed largely of DM and form very early in the universe ($200 \leq z \leq 1000$)~\citep{ricotti2009,bringmann2012,bertschinger1985}. The local environment can prove fruitful as it is open to investigation via two complimentary and independent probes. These being indirect DM searches~\citep{bringmann2012} and microlensing surveys hunting unknown dark structures~\citep{li2012}. The power of both probes is confined locally due to the fact that experiments like Gaia cannot accurately reconstruct microlens masses more a few 10's of kpc from the solar system~\citep{gaia2001}, and that DM-induced emission from such small structures would be extremely faint. Investigations of the existence of these objects via indirect DM detection have only produced limits on their abundance within the Milky-Way that depend upon the annihilation cross-section matching the relic value. Thus, these constraints have not been examined with the use of an experimental upper-limit on the DM annihilation cross-section~\citep{bringmann2012}.

UCMHs make a particularly attractive target for investigation of micro-structure as their formation has been argued for via a variety of mechanisms. Allowing them to act as probes of a variety of possible physics. These range from cosmic strings and other topological defects~\citep{silk1993}, to small-scale effects of inflation~\citep{aslanyan2015}, or phase-transitions in the early universe~\citep{josan2010}. The smaller density contrast of the necessary primordial density perturbations means that the population of these objects is far less severely constrained by cosmic power spectra than that of primordial black holes~\citep{blais2003,josan2010,bringmann2002}. Given the broad range of available formation scenarios, and the fact that their high central density means they are very likely to survive until the present epoch~\citep{berezinsky2006,berezinsky2006,bringmann2012}, the possibility of these objects is one to take seriously.

As such, the abundance of these UCMHs is able to probe the non-Gaussianity of the primordial density perturbation power-spectrum~\citep{clark2015} as well as general small-scale detail~\citep{li2012}, distant cosmic epochs like that of electron-positron annihilation~\citep{josan2010}, as well as details of inflation~\citep{aslanyan2015}, and can also have measurable effects on pulsar timing arrays~\citep{clark2015}. These properties place these objects at the fore-front of the determination of vital details about the completeness of the concordance cosmological scenario.

Additionally, because UCMHs are very dense, they may offer unique opportunities to observe DM annihilation/decay signals in nearly `naked' DM halos as well as making them excellent targets for lensing observations. The DM-dominated nature of these objects arises as dense core is not vulnerable to tidal stripping if the ratio of the core radius to that of the whole halo is $\lesssim 10^{-3}$~\citep{bringmann2012}. But, any accreted baryonic matter would likely be removed in this manner~\citep{ricotti2009,bertschinger1985}. This implies that little in the way of purely baryonic emission should be expected from such structures.
In the context of annihilating DM it has also recently been suggested that, if sufficiently abundant, compact DM halos may contribute significantly to the heating of interstellar gas within galaxies~\citep{clark2016}.
Furthermore, a link has been recently conjectured between the Earth encountering such objects and catastrophic events occurring on terrestrial planets within the solar system~\citep{dino1,dino2,dino3}. Such as mass extinction events caused by increased thermal activity in the Earth's core due to WIMP annihilation resulting from the planetary accretion of WIMPs from within the highly dense centre of the UCMH.

However, in order examine the multi-faceted consequences of these objects, their abundance must be determined, or at least limited. This proves to be difficult, as the emissions of such small objects from DM annihilation/decay are expected to be faint, and the use of independent limits on DM annihilation cross-sections is ineffectual as, when the search is conducted exclusively through indirect detection mechanisms, setting an upper-limit on UCMH abundance requires a lower bound on the cross-section~\citep{bringmann2012}. This is an obviously problematic scenario, as despite the possibility of using the canonical thermal relic cross-section~\citep{jungman1996} as a lower bound, candidates for the formation of such small structure (like WIMPs) are being pushed into phase-space territory below this cross-section by multiple sets of data and for a variety of DM masses~\citep{planck2014,gsp2015,Fermidwarves2015}. Thus, in order to examine these sorts of DM models through the attractive lens of local UCMHs, a method of tying experimental bounds on WIMP annihilation to those on the abundance of ultra-compact structures is necessary.

To obviate this difficulty, we propose a means of hunting compact dark objects by combining local microlensing measures with indirect DM detection techniques. This will be shown to by-pass the problem of linking experimental bounds on UCMH abundance and WIMP annihilation by setting a search region where microlensing observations could accurately determine the presence and mass of previously unknown compact structures. The region in question will encompass the densest parts of the Milky-Way and thus the regions of highest UCMH probability~\citep{ricotti2009,bringmann2012}, this is necessary to ensure the reliability of the extrapolation of a null detection to the rest of the Milky-Way. The lensing search allows us to set a definite radius on the indirect probe provided we require a correlation between unknown dark microlensing objects and purported indirect DM signatures. This solution is effective because it removes the degeneracy between annihilation cross-section and UCMH limits~\citep{bringmann2012} by no longer using the cross-section and experimental sensitivity of the indirect search to set the maximal radius probed for these objects. This comes at a price of somewhat weaker possible limits (as the search is less exhaustive), but it ensures the limits on UCMH abundance are indeed meaningful and robustly linked to WIMP annihilation limits. Another point in the favour of this search methodology is that it would allow for an independent corroboration of the mass of all candidates for unknown compact objects as a consistency check on indirect detection claims. Additionally, distributional data for candidate objects will be necessary for extending abundance limits beyond the null scenario.

For concreteness we will utilise the fact that so far no UCMH object has been officially observed to set limits on the abundance of these objects by extrapolating from a lack of indirect detections within a radius that can be effectively probed by Gaia microlensing studies~\citep{gaia2001,gaia-docs}. This will be done using the sensitivity of the Fermi-LAT telescope~\citep{fermi-docs}\footnote{\url{http://www.slac.stanford.edu/exp/glast/groups/canda/lat_Performance.htm}} and the Fermi mission limits on WIMP annihilation~\citep{Fermidwarves2015}. The radius set by Gaia proves very similar to the maximum radius that can be probed by Fermi-LAT over a wide range of UCMH masses when Fermi-LAT's own limits on annihilation cross-sections are used. Such results will then be contrasted with the prospective limits from the Square Kilometre Array (SKA)~\citep{ska2012} as well as the Cherenkov Telescope Array (CTA)~\citep{funk-cta2013} should there by a null result on UCMH identification with these telescopes within the region that Gaia could accurately characterise candidates. The limits on UCMH abundance, thus derived, will then be used to produce limits on the behaviour of very small scales in the power spectrum of primordial density perturbations, following a methodology similar to \citep{aslanyan2015}. The use of a finite search radius defined independent of our indirect detection experiments allows us to also make direct comparison of the prospects for each of the SKA, Fermi-LAT, and the CTA.

Finally, we want to explore the possible impact of UCMHs on the very local scale of our solar system. In this context,  we explore the hypothesis of volcanogenic DM whereby capture of WIMPs by the Earth during transit through very dense DM mini-halos produces an up-surge in volcanic activity due to heat being deposited within the Earth's core via DM annihilation/decay. We will show that there are significant doubts about the magnitude of this effect, by using the halo model of \citep{ricotti2009,bringmann2012}, at variance with previous work on this topic which used a UCMH model derived by consideration of UCMH seeding by cosmic strings~\citep{silk1993}. This latter model produces UCMHs with considerably larger central cores than the more general and cosmologically consistent formation scenario derived by \citep{ricotti2009,bringmann2012}.
In fact, the effect of the more general halo model is to reduce the significance of the predicted planetary core heating by two orders of magnitude. In addition to this, we will show that taking into account the elemental composition of the Earth, and the LUX limits on  WIMP-nucleon cross-section~\citep{lux2013} reduces the significance of DM-induced heating of the Earth's core even further. Later results on WIMP-nucleon scattering~\citep{lux2016}, or arXiv: 1705.06655, reduce significance by an additional order of magnitude. Moreover, we also demonstrate that, for UCMHs large enough to induce significant core heating in Earth, the time-scale for solar system encounters with UCMHs is largely incompatible with the suggested 30 Myr extinction cyle on Earth~\citep{dino2}. However, we note that the heat flow within the Martian core is much lower than that within Earth, and thus we will determine that considerably more significant heat-flow can be generated within the Martian core by DM capture. We use this to motivate the ``de-gassing" scenario presented by \citep{sandu2012}, where a significant increase in volcanism/mantle melt-zone activity leads to the shutting down of the Martian geodynamo and thus the loss of atmosphere~\citep{maven} and life-supporting conditions on the planet. Importantly, we will also examine the regions of the parameter space that can provide sufficient encounter frequencies to support the Martian de-gassing. We will show that models with WIMP masses below 100 GeV and UCMH masses below 10$^2$ M$_{\odot}$ are not constrained by Fermi-LAT, and, those with WIMPs annihilating via quarks will not be further constrained by SKA/CTA.

This paper is structured as follows: the properties of UCMHs are detailed in section~\ref{sec:ucmh}, calculation of their abundance is given in section~\ref{sec:abundance}, the consequences of DM annihilations are described in section~\ref{sec:dmann}, decaying DM in section~\ref{sec:dmdecay}, while the relative multi-frequency emissions are explored in section~\ref{sec:emm}. The results obtained for the detectability of UCMHs by various telescopes are discussed in section~\ref{sec:detect}, and the constraints on the UCMH abundance are discussed in section~\ref{sec:large}. The cosmological constraints provided by UCMHs are presented in section~\ref{sec:cosmo}.
Solar system constraints on the heating of the Earth's core and the consequences for mass extinction, as well as the constraints of the Mars core heating and the magnetic field loss of the planet  are discussed in section~\ref{sec:local}.
Finally, our discussion and conclusions are presented in section~\ref{sec:conclusions}.

\section{Ultra-Compact Mini-Halos}
\label{sec:ucmh}

As discussed in \citep{ricotti2009}, over-dense regions dominated by DM with a density contrast as small as $\delta \sim 10^{-3}$ during the radiation dominated epoch, could later collapse into very compact halo-type objects. This was originally derived by an order of magnitude argument, but in \citep{bringmann2012} a detailed calculation demonstrates that this is indeed the case. Moreover, it was demonstrated that these density perturbations associated to UCMHs could collapse in the interval $200 \lesssim z \lesssim 1000$ from similar initial over-density levels. The early formation times of these perturbations mean that they do not initially contain any baryonic matter or radiation, as both of these free-stream out of the over-dense region.  However, baryonic matter can be later accreted after the time of decoupling~\citep{bringmann2012}. Since the necessary density perturbations required to form UCMHs are of much lower amplitude than those required to form primordial black holes~\citep{blais2003,josan2010,bringmann2002}, the corresponding  distortions of the perturbation power-spectrum are substantially less severe on small scales, and this fact produces less severe constraints on their abundance.

In order to set constraints on the UCMH abundance we have to describe their formation process and their resulting structure.\\
The initial mass of DM (assumed to be in the form of WIMPs $\chi$) contained in an over-dense region of co-moving size $R$ is given by
\begin{equation}
M_i = \frac{H_0^2}{2 G} \Omega_{\chi} R^3 \; ,
\end{equation}
where $H_0$ is the value of the redshift-zero Hubble constant, $G$ is the gravitational constant, and $\Omega_{\chi}$ is the energy density fraction of WIMP DM in the universe. \\
The mass $M_i$ can be re-expressed in solar mass units as
\begin{equation}
M_i = 1.3 \times 10^{11} \left( \frac{\Omega_{\chi}h^2}{0.112}\right) \left(\frac{R}{\mbox{Mpc}}\right)^3 \;\mbox{M}_{\odot} \; .
\end{equation}
The simple dependencies of this expression are entirely due to the fact that the initial perturbation is formed entirely of DM particles, and none of the relativistic degrees of freedom enter into the formation of the seed mass, as both matter and radiation will free-stream out of the over-dense region given its formation in the radiation dominated epoch.

This initial seed mass $M_i$ will then be able to accrete further DM particles after the epoch of matter-radiation equality at $z_{eq}$, as well as baryonic matter after the decoupling epoch. If accretion is assumed to take place from a smooth, unbound cosmological DM background, then the UCMH mass will grow as
\begin{equation}
M_{UCMH} = M_i \frac{1+z_{eq}}{1+z} \; .
\end{equation}
An accretion process of this nature will of course become inefficient once hierarchical structure formation has reached a point where there is sufficient dynamical friction between the newly forming structures and the UCMHs. This is due to the now significant larger heirarchical structure tidally stripping any newly accreted matter from the periphery of the UCMH.
The conservative choice made in \citep{bringmann2012} is that this point is reached around $z \sim 10$; we will similarly adopt this limit on the growth of UCMH perturbations.

Since, after kinetic decoupling of the WIMP from the Standard Model particles (which occurs after the WIMP abundance has frozen out~\citep{bringmann2009}), UCMHs form entirely via radial in-fall~\citep{ricotti2009,ricotti2008,bringmann2009} (as their is no significant scattering of WIMPs by other matter) this leads to a density profile of the form
\begin{equation}
\rho_{UCMH} (z) = \frac{3 f_{\chi} M_{UCMH} (z)}{16 \pi R_{UCMH}^{\frac{3}{4}}(z) r^{\frac{9}{4}}} \; , \label{eq:rho}
\end{equation}
where $f_{\chi}$ is the fraction of matter in the form of DM, and $R_{UCMH}(z)$ is the effective radius of the UCMH at redshift $z$. This effective radius is given by the following expression derived from numerical simulations~\citep{ricotti2007,ricotti2008}
\begin{equation}
R_{UCMH} (z) = 0.019 \left(\frac{1000}{1+z}\right) \left( \frac{M_{UCMH}(z)}{\mbox{M}_{\odot}}\right)^{\frac{1}{3}} \; . \label{eq:size}
\end{equation}
The extremely steep halo density profile $\sim r^{-9/4}$ is a consequence of the spherically symmetric collapse, as derived in \citep{fillmore1984,bertschinger1985} and later demonstrated by numerical simulations~\citep{vogelsberger2009,ludlow2010}.\\
Since this radial-infall profile cannot be valid down to $r = 0$, a cut-off radius $r_{min}$ is  established, below which the density will be set constant, as later accretion will preferentially influence the outer parts of the halo~\citep{ricotti2009,bertschinger1985}.\\
Thus, the DM density within the UCMH will be limited by
\begin{equation}
\rho (r \leq r_{min}) = \rho(r_{min}) \; ,
\end{equation}
where $r_{min}$ is defined by the expression from \citep{bringmann2012}
\begin{equation}
r_{min} = 2.9\times 10^{-7} \left(\frac{1000}{1 + z_c}\right)^{2.43} \left( \frac{M_{UCMH}(0)}{\mbox{M}_{\odot}}\right)^{-0.06} R_{UCMH}(0) \; ,
\end{equation}
where $z_c$ is the redshift at which the UCMH collapsed.
However, there is a second physical process that may limit the cuspiness of the halo. That of DM annihilation.
The constant value of the density profile within a second radius $r_{cut}$ is established by noting that UCMHs  will have their density profile affected by the annihilation of DM, which is particularly efficient within the central cusp region. This will result in a maximal density within the cusp producing a flat halo density profile at $r \leq r_{cut}$.  Since this requires knowledge of the maximal annihilation-limited density, we will make use of the expression
\begin{equation}
\rho_{max} = \frac{m_{\chi}}{\langle \sigma V\rangle (t-t_i)} \; ,
\label{eq:rhomax}
\end{equation}
with
\begin{equation}
\rho (r \leq r_{cut}) = \rho_{max} \, \label{eq:rhomax2}
\end{equation}
where $m_{\chi}$ is the mass of the WIMP with a thermally-averaged annihilation cross section of $\langle \sigma V\rangle$.\\
For UCMHs at the present epoch $t$ is 13.799 Gyr~\citep{planck2014} and $t_i$ is taken to be equal to $t_{eq}$ following the arguments given in \citep{scott2009,wright2006}. Thus, if $r_{cut} < r_{min}$ the cut-off radius and maximal density is determined by $r_{min}$. But, if $r_{cut} > r_{min}$, then the core density of the UCMH will be limited by Eq.(\ref{eq:rhomax2}), and will automatically satisfy the $r_{min}$ requirement as well. However, in the case of decaying DM, we will only use the $r_{min}$ flattening. This may be optimistic but will largely not affect the conclusions drawn here due to the weakness of the possible constraints on decaying UCMHs.

\section{Abundance of Ultra-Compact Halos}
\label{sec:abundance}

In order to derive constraints on the possible population of UCMHs we need to calculate their abundance. The proper way to proceed to determine the population statistics and abundance of UCMHs within a structure like a galaxy would be to calculate a mass function for these objects. However, this function would depend strongly on the formation mechanism assumed for the production of UCMHs. Thus, in order remain model independent and avoid such assumptions, we will proceed in the same manner as \citep{bringmann2012} and determine the abundance of each mass of UCMH separately based on a model independent probability that some overdense region will collapse to form a UCMH. This method can be model independent as the index of the power spectrum $n_s$ of primordial density perturbations is modified at each scale by a local, scale-dependent value of the running of the index $\alpha_s (k) = \frac{d n_s}{d \log{k}}$.

For Gaussian distributed perturbations, the probability that an overdense region of co-moving radius $R$ collapses into a UCMH is given by the expression
\begin{equation}
\beta_U(R) = \frac{1}{\sqrt{2 \pi} \sigma_{\chi,H}(R)}\int_{\delta_{min}}^{\delta_{max}} \exp{\left[ -\frac{\delta^2}{2 \sigma_{\chi,H}^2(R)}\right]}\, d\delta \; ,
\end{equation}
where $\sigma_{\chi,H}(R)$ is the r.m.s. density perturbation (or DM mass variance function) taken at the point of horizon entry for the density perturbation, $\delta_{min} \sim 10^{-3}$, and $\delta_{max} \sim \frac{1}{3}$~\citep{ricotti2009,bringmann2012,josan2010}. This means that the fraction of DM in the form of UCMHs is given by
\begin{equation}
\Omega_{UCMH} = \Omega_{\chi}\frac{M_{UCMH}(0)}{M_i} \beta_U(R_{UCMH}(0)) \; .
\end{equation}
We note that there is no accounting, in this expression, for the possibility of UCMHs being destroyed by tidal forces and mergers.
This is due to the early formation time of these ultra compact structures, which ensures that they collapse into extremely over-dense objects by the time general structure formation has commenced. In particular, the core density of a UCMH of one solar mass, assuming a 1 TeV WIMP with canonical relic cross-section, is a factor of $10^{14}$ greater than the average matter density when $z \lesssim 30$.
Additionally, the density profile of these halos is so steep as to almost ensure that the over-dense UCMH survives tidal effects, as core-to-effective radii ratios $\frac{r_{cut}(z)}{R_{UCMH}(z)}$ of $\lesssim 10^{-3}$ indicate a probability of the halo enduring into the present epoch that is nearly equal to one according to numerical simulations~\citep{berezinsky2006,berezinsky2008}.
Moreover, DM emissions are mostly dependent on the central core of the halo, and this is sufficiently dense as to prevent tidal stripping effects having an impact on the indirect detection of UCMHs as the tidally stripped matter will come from the outer parts of the halo in this case~\citep{ricotti2009,bertschinger1985}. It must be emphasised that the ratio of core-to-effective radii is $\lesssim 10^{-4}$ for all UCMHs formed before $z = 200$, for almost any given mass. Another important consequence of the extreme over-density is that the UCMH spatial distribution is expected to track that of the smooth DM halo. This is because the compact halos are too dense to be biased towards larger radii, unlike other sub-halo DM clumps that are vulnerable to tidal disruption~\citep{springel2008}.

In order to evaluate $\beta_U$ we need the expression for the mass variance function
\begin{equation}
\sigma^2_{\chi,H}(R) = A_{\chi}^2(k_R) \delta_{\chi,H}^2 (k_R) \; ,
\end{equation}
where $k_R$ is the wave-number corresponding to the perturbation of radius $R$, and
\begin{equation}
\begin{aligned}
A_{\chi}^2 (k_R) = \frac{9}{16} \int_0^{\infty} dx \; x^{n_s + 2 +\alpha_s \log{\left(\frac{x k_R}{k_*}\right)}} \left(\frac{x k_R}{k_*}\right)^{\alpha_s \log{\left(\frac{x}{k_*}\right)}} \\ \times W_{TH}^2 (x) \frac{T_{\chi}^2(x/\sqrt{3})}{T_r^2(1/\sqrt{3})} \; . \label{eq:pspec}
\end{aligned} 
\end{equation}
The running of $n_S$ is given by $\alpha_s = \td{n_s}{\log{k}}{}$, and we can self-consistently include a running of the running parameter $\beta_s = \td{\alpha_s}{\log{k}}{}$. The quantities $T_{\chi}$ and $T_r$ are the DM and radiation transfer functions given in \citep{bringmann2012}, and $W_{TH}(x)$ is the Fourier transform of the top-hat window function.
We note that the function $\beta_U$, as well as any derived constraints on $\alpha_s$ and $\beta_s$, will be local in $k$. This is due to the fact that, since no UCMH population has yet been found, each $k$ is constrained independently~\citep{bringmann2012}. The importance of this point is that it implies that the constraints on the power spectrum cannot be globally extrapolated, they are relevant only to a given wave-number. We follow this approach in order to avoid assumptions about the nature of the mass distribution of the hypothetical UCMH population.\\
The amplitude $\delta_{\chi,H}$ is defined to match CMB normalisation~\citep{bringmann2012,aslanyan2015}
\begin{equation}
\delta^2_{\chi,H}(k_R) = \delta_0^2 \left(\frac{k}{k_*}\right)^{n_s + -1 +\alpha_s \log{\left(\frac{k}{k_*}\right)}} \; , \label{eq:delta-chi}
\end{equation}
with
\begin{equation}
\delta_0^2 = \frac{4.58 \times 10^{-10}}{1+0.53 r_{ts}} \exp(2 (1 - n_s)(-1.24 + 1.04 r_{ts})) \; ,
\end{equation}
where $r_{ts}$ is tensor-to-scalar ratio, and $k_* = 0.05$ Mpc$^{-1}$.

The fraction $f$ of DM within the Milky-Way that is found in UCMHs of mass $M_{UCMH}(0)$ is then defined as
\begin{equation}
f = \frac{M^{tot}_{UCMH}}{f_{\chi} M_MW} \; ,
\end{equation}
where $M^{tot}_{UCMH}$ is the total mass of UCMHs of mass $M_{UCMH}(0)$, and $f_{\chi} M_{MW}$ is the mass of DM in the Milky-Way.
The limiting expression for the maximum fraction of DM found in UCMHs from their non-detection has been derived in~\citep{bringmann2012} as
\begin{equation}
f_{max} = \left(\frac{F_{min}(\nu)}{F(\nu,d)}\right)^{\frac{3}{2}} f_{\chi} \frac{3 M_{UCMH}(0)}{4\pi \rho_{\chi}(d) d^3}\ln{\left(1 - \frac{\sigma_{obs}}{\sigma_{exp}}\right)} \; ,
\label{eq:fmax}
\end{equation}
where $\rho_{\chi}(d)$ is the DM density at a distance $d$, $F(\nu,d)$ is the expected flux from a UCMH at $d$, $d_{obs}$ is the maximum observed radius from the Earth, $F_{min}(\nu) \equiv F(\nu,d_{obs})$ is the minimal expected flux from a UCMH within $d_{obs}$, $\sigma_{obs}$ is the confidence level for which $f \leq f_{max}$, and $\sigma_{exp}$ is the confidence level at which we should then see some flux from a UCMH candidate, located at $d_{obs}$, which exceeds the sensitivity threshold of the experiment in question. We note here that $d_{obs}$ is not defined explicitly by the sensitivity of a given experiment and the WIMP annihilation cross-section. We motivate this choice on the basis of our dual observation methodology. The region within which we will search for indirect DM emissions from UCMHs will be dictated by what region can be accurately probed by microlensing studies with Gaia, as, without a suitably characterised candidate UCMH object, purported indirect DM signatures are far from robust. The radius explored will be a few 10's of kpc at most~\citep{gaia2001} for a 50\% accuracy level in mass reconstruction. We will thus confine our search region to a radius of 10 kpc, noting that we include the most probable regions for UCMH detection (as their distribution is expected to trace that of the smooth DM component).

It is also important to note
that Eq.~(\ref{eq:fmax}) was derived under the assumption that all UCMHs have the same mass, as no UCMHs have yet been observed. As a result of this, and in order to avoid dependence on the model generating the UCMH seeds, we will thus constrain UCMHs of each given mass independently.

\section{Ultra-compact Halos in the Solar System}
\label{sec:capture}

The capacity of UCMHs to survive un-altered from their early formation epoch up to the local universe allows these structures to pervade our local environment such as our galaxy and its inner regions, like the solar system. Thus, it is justified to explore the impact of such UCMHs on planetary system structure and evolution. This will build on the previous section as the abundance of UCMHs within the Milky-Way will be vital to determine how often it would be possible for the solar system to encounter such halos. In order to determine the periodicity $\tau$ with which the solar system will encounter UCMHs we will make use of the formula from~\citep{collar1996}
\begin{equation}
\tau = \left(f \frac{\rho_{local}}{M_{UCMH}(0)} \sigma_v (0) \sigma_A  \right)^{-1} \; ,
\end{equation}
where $\rho_{local} \approx 0.3$ GeV cm$^{-3}$ is the local DM density, $\sigma_v$ is the DM velocity dispersion, and $\sigma_A$ is the cross-sectional area of the UCMH (we suppress this area by a factor of $\bar{\rho}/\rho_{max}$ to reflect the steepness of the halo profile). Making use of Eqs.~(\ref{eq:size}), assuming a spherical halo profile for the UCMH, and employing an approximate value of $\sigma_v \approx 300$ km s$^{-1}$~\citep{collar1996}, 
we find that when $f \sim 0.1$, $\tau \sim \mathcal{O}(10^{-1})$ Gyr (in agreement with previous work~\citep{collar1996} at the corresponding halo mass). However, if we use $f \sim 10^{-7}$ we find that $\tau \sim \mathcal{O}(10^5)$ Gyr. This indicates that multiple halo encounters are only probable with a large/poorly-constrained UCMH abundance. This point will be discussed in more detail in Section~\ref{sec:mag}.

If WIMP DM annihilates/decays, then the capture of WIMPs within the core of a planet might lead to an increase of the heating inside this region or of the adjacent mantle. It has been argued that a sufficiently dense clump of DM could raise the temperature of the Earth's core by several hundred degrees Kelvin~\citep{dino1,dino3}. This is the hypothesis of ``volcanogenic" DM: i.e., encounters with dense DM clumps cause volcano-induced mass extinction events on a $\sim$ 30 Myr cycle~\citep{dino1,dino3}. However, as observed above, the periodicity of the UCMH encounters is highly sensitive to the abundance $f$. In order to accommodate sufficiently large $f$ values, the UCMH mass ranges allowed by the Fermi-LAT data for a 1 TeV WIMP with the canonical cross-section would have to be $M_{UCMH}(0) \leq 10^{-7}$ M$_{\odot}$ or $M_{UCMH}(0) \geq 10^{11}$ M$_{\odot}$~\citep{bringmann2012}. However, using galactic diffuse emission data from the Fermi mission, the same authors place a restriction of $f < 10^{-3}$ for all UCMH masses between $10^{12}$ and $10^{-8}$ M$_{\odot}$ for the aforementioned WIMP model. This implies that the most optimistic periodicity we can achieve will be $\tau \sim \mathcal{O}(10)$ Gyr, three orders of magnitude above the conjectured mass extinction periodicity. This already seems to deal fatal damage to a notion of volcanogenic DM being correlated with mass extinctions but we will further investigate what happens if such an object was encountered. Moreover, we will pay particular attention to how generalising the WIMP mass considered affects these conclusions.\\
A related hypothesis has been formulated that a smooth DM distribution in a disk on the galactic-plane would increase the probability of large meteor impacts that can trigger extinction events~\citep{dino2} as the solar system orbits through the disk. Indeed there is evidence of a periodic cycle of increased cratering on Earth that appears to be non-random in origin~\citep{raup1984,bailer2011} with a period very similar to that with which the solar system passes through the galactic plane~\citep{rampino1984,schwartz1984}. The suggested mechanism for this increased cratering is that the Oort cloud is perturbed by the passage through the denser galactic disk, perturbations of this nature have been linked with comet showers entering the inner regions of the solar system~\citep{rampino1984,schwartz1984}. Molecular clouds were originally suggested to explain the cratering periodicity, but have been shown to be spread too far from the galactic plane~\citep{thaddeus1985}, leaving the authors in \citep{dino2} to propose DM as providing sufficient density in the galactic plane to perturb the Oort cloud.\\
A third hypothesis is that of mass extinction induced by DM scattering off of atoms with the DNA of living creatures, essentially a mass-cancer outbreak induced by the passage of the solar system through a highly dense DM clump~\citep{collar1996}. It has been argued, however, that such an explanation of mass extinctions is at odds with geological records~\citep{dino1}.\\
Here we will specifically address the volcanogenic DM hypothesis, as it has direct bearing on annihilating/decaying DM in UCMHs that we discuss in this paper.

The first ingredient in determining the heating of a planetary core via WIMP annihilation/decay is their capture rate by the planet itself, which is given by~\citep{krauss1986,dino3}
\begin{equation}
N_E = 4.7 \times 10^{17} \; \mbox{s}^{-1} \; 3 ab \rho_{0.3} v_{300}^{-3} \sigma_{N,32} \left( 1 + \frac{m_{\chi}^2}{m_N^2}\right)^{-1} \frac{M_P}{M_{Earth}} \; ,
\end{equation}
where $M_P$ is the planet mass, and $M_{Earth}$ is that of the Earth, the quantity $ab \sim 0.34$ is derived from \citep{dino1}, $\rho_{0.3}$ is the DM density normalised to $0.3$ GeV cm$^{-3}$, $v_{300}$ is the DM velocity normalised to 300 km s$^{-1}$, $\sigma_{N,32}$ is the WIMP-nucleus scattering cross-section normalised to $10^{32}$ cm$^2$, $m_{\chi}$ is the WIMP mass, and $m_N$ is nuclear mass.

An alternative, but more complex expression, was used in \citep{gould1987}:
\begin{equation}
N_E = 4.0 \times 10^{16} \; \mbox{s}^{-1} \; \frac{M_P}{M_{Earth}} \rho_{0.4} \frac{\mu}{\mu^2_+} Q^2 f_E \left\langle \hat{\phi} \zeta_1(A) \right\rangle \; , \label{eq:gould}
\end{equation}
where $\rho_{0.4}$ is the DM density normalised to $0.4$ GeV cm$^{-3}$, $f_{E}$ is the core mass fraction due to the element $E$ under consideration, the angular brackets denoted averaging over the planet's core, and the $Q$ function is defined by
\begin{equation}
\begin{aligned}
Q \approx N - 0.124 Z \; ,
\end{aligned}
\end{equation}
where $N$ is number of neutrons in the nucleus and Z is the corresponding proton number.
Then the remaining functions are defined as follows
\begin{equation}
\begin{aligned}
\mu &= \frac{m_{\chi}}{m_N} \; , \\
\mu_+ &= \frac{1}{2}(\mu + 1) \; , \\
\mu_- &= \frac{1}{2}(\mu - 1) \; , \\
A &= \sqrt{\frac{3\mu}{2 \mu_-^2}}\frac{v}{\bar{v}} \; , \\
A_+ &= A + 1  \; , \\
A_- &= A - 1 \; , \\
\hat{\phi} &= \frac{v^2}{v_{esc}^2} \; , \\
\chi(a,b) &= \frac{\sqrt{\pi}}{2}\left( \mbox{erf}(b) - \mbox{erf}(a) \right) \; , \\
\zeta_1(A) &= \frac{1}{2 A^2}\Bigg[ \left( A_+A_- - \frac{1}{2}\right)\left( \chi(-1,1) - \chi(A_-,A_+) \right) \\ & + \left(\frac{1}{2}A_+ \mbox{e}^{-A_-^2} + \frac{1}{2}A_-\mbox{e}^{-A_+^2}\right) - \mbox{e}^{-2} \Bigg]  \; .
\end{aligned}
\end{equation}
Here $v$ and $\bar{v}$ represent the DM velocity and its average respectively, while $v_{esc}$ is the escape velocity for a DM particle in a given capture scenario.\\
For the Earth's core \citep{gould1987} gives $A \sim \frac{\sqrt{\mu}}{17 \mu_-}$ with $\langle \hat{\phi}\rangle = 1.6$. These factors are employed to better understand capture damping due to DM velocity, as well as capture enhancement due to resonant scattering. We will look here at two models: one where we assume that the entire planet is composed of iron nuclei to discuss a best-case scenario; in the second model we will use a more accurate break-down of the chemical composition~\citep{earthCompo} (we will assume that the Earth and Mars have similar composition fractions).

The accretion rate calculation from Eq.~(\ref{eq:gould}) is normalised to the case where the WIMP-nucleus cross-section is given by
\begin{equation}
\begin{aligned}
\sigma & = \frac{\mu}{\mu_+^2} Q^2 \frac{m_N m_{\chi}}{(\mbox{GeV})^2} \times 5.2 \times 10^{-40} \; \mbox{cm}^2 \; ,
\end{aligned}
\end{equation}
from which we can deduce that the WIMP-nucleon cross-section is assumed to follow the form
\begin{equation}
\begin{aligned}
\sigma_N & = 4 \left(\frac{m_n}{m_{\chi}}\right) \left( \frac{m_n}{m_{\chi}} + 1 \right)^2 \left(\frac{1}{0.124}\right)^2 \left(\frac{m_{\chi}}{\mbox{GeV}}\right) \left(\frac{\mbox{GeV}}{m_n}\right)\times 5.2 \times 10^{-40} \; \mbox{cm}^2 \; ,
\end{aligned}
\end{equation}
where $m_n$ is nucleon mass.
In addition to this, we will also explore the case where the capture rate is renormalised to WIMP-nucleon cross-section limits from the LUX experiments $\sigma_{N,LUX}(m_{\chi})$~\citep{lux2013} (using \citep{lux2016,xenon1t} changes results by less than an order of magnitude)
\begin{equation}
\sigma_N = 4 \left(\frac{m_n}{m_{\chi}}\right) \left( \frac{m_n}{m_{\chi}} + 1 \right)^2 \left(\frac{1}{0.124}\right)^2 \left(\frac{\mbox{GeV}}{m_n}\right) \sigma_{N,LUX}(m_{\chi}) \; .
\end{equation}

The power produced by the annihilation/decay of captured WIMPs is then expressed as
\begin{equation}
P_E = e N_E m_{\chi} \; ,
\end{equation}
where $e$ is the fraction of annihilations/decays that lead to the heating of the target planet's core. We follow \citep{dino3} in the assumption that it takes $1.56 \times 10^{27}$ ($2.74 \times 10^{25}$) J to raise the core temperature of the Earth (Mars) by $1$ K and that the non-DM power produced in Earth's core is $4\times 10^{13}$ W~\citep{mack2007}. Additionally, we note that the power required to sustain the geodynamo of Mars is taken to be $3 \times 10^{11}$ W~\citep{sandu2012}. We shall take $e = 1$ to present the very maximum possible effects of transit through a dense UCMH, with the planet's trajectory following the diameter line of the compact halo at $220$ km s$^{-1}$. Note that we consider only passage through the region of the UCMH between $\rho_{max}$ and $10^{-5}\rho_{max}$ in order to capture the significant effects of the halo (we calculate crossing times appropriately accounting for this).

The DM density appears as a prominent part of both the above capture rate formulae, as such we will use the full profile from \citep{bringmann2012}. We stress that previous work~\citep{dino1,dino3} has used a model where the UCMH is treated as having a wide flat core and negligible outer regions of the halo $R_{UCMH} \approx r_{core}$. This approximation is drawn from \citep{silk1993}, which considered UCMH forming over-densities that arise from cosmic strings or textures in the early universe. The premise of the formation of these UCMHs leads to them having substantially larger core regions than the very general formalism from \citep{bringmann2012}, as well as densities set uniformly at the DM density at the time of matter-radiation equality. The use of the more general profile will prove to be highly significant.

\section{Dark Matter Annihilation}
\label{sec:dmann}

For the annihilation of a pair of WIMPs we will consider the source function for particle species $i$ to be given as a function of product energy $E$, and radius within the UCMH $r$, by
\begin{equation}
Q_i (r,E) = \langle \sigma V\rangle \sum\limits_{f}^{} \td{N^f_i}{E}{} B_f \left(\frac{\rho_{\chi}(r)}{m_{\chi}}\right)^2 \; ,
\end{equation}
where $\langle \sigma V\rangle$ is the non-relativistic velocity-averaged annihilation cross-section at $0$ K, $B_f$ is the branching fraction for intermediate state $f$, $\td{N^f_i}{E}{}$ is the differential $i$ yield of the $f$ channel, and $\left(\frac{\rho_{\chi}(r)}{m_{\chi}}\right)^2$ is the WIMP pair number density.

The functions $\td{N^f_i}{E}{}$ will be sourced from the Pythia~\citep{pythia} routines within the DarkSUSY~\citep{darkSUSY} package as well as \citep{ppdmcb1,ppdmcb2}. We will follow the standard practice of studying each annihilation channel $f$ independently, assuming $B_f = 1$ for each separate case.

\section{Dark Matter Decay}
\label{sec:dmdecay}

For the decay of WIMPs we will consider the source function for particle species $i$ to be given as a function of product energy $E$, and radius within the UCMH $r$, by
\begin{equation}
Q_i (r,E) = \Gamma \sum\limits_{f}^{} \td{N^f_i}{E}{} B_f \frac{\rho_{\chi}(r)}{m_{\chi}} \; ,
\end{equation}
where $\Gamma$ is the decay rate, $B_f$ is the branching fraction for intermediate state $f$, $\td{N^f_i}{E}{}$ is the differential $i$ yield of the $f$ channel, and $\frac{\rho_{\chi}(r)}{m_{\chi}}$ is the WIMP number density.

The functions $\td{N^f_i}{E}{}$ will be sourced from \citep{ppdmcb1,ppdmcb2}. We will follow the standard practice of studying each decay channel $f$ independently, assuming $B_f = 1$ for each separate case.

\subsection{Dark Matter emissions}
\label{sec:emm}
The major products of DM annihilation/decay within the UCMHs that we will be interested in are electrons and prompt $\gamma$-ray emission. Thus, the processes we will account for are synchrotron radiation, inverse-Compton scattering of CMB photons, bremsstrahlung, and prompt $\gamma$-ray production by $\pi^0$ decay.

The average power of the synchrotron radiation at observed frequency $\nu$ emitted by an electron with energy $E$ in a magnetic field with amplitude $B$ is given by~\citep{longair1994}
\begin{equation}
P_{synch} (\nu,E,r,z) = \int_0^\pi d\theta \, \frac{\sin{\theta}^2}{2}2\pi \sqrt{3} r_e m_e c \nu_g F_{synch}\left(\frac{\kappa}{\sin{\theta}}\right) \; ,
\label{eq:power}
\end{equation}
where $m_e$ is the electron mass, $\nu_g = \frac{e B}{2\pi m_e c}$ is the non-relativistic gyro-frequency, $r_e = \frac{e^2}{m_e c^2}$ is the classical electron radius, and the quantities $\kappa$ and $F_{synch}$ are defined as
\begin{equation}
\kappa = \frac{2\nu (1+z)}{3\nu_g \gamma^2}\left[1 +\left(\frac{\gamma \nu_p}{\nu (1+z)}\right)^2\right]^{\frac{3}{2}} \; ,
\end{equation}
with $\nu_p \propto \sqrt{n_e}$, and
\begin{equation}
F_{synch}(x) = x \int_x^{\infty} dy \, K_{5/3}(y) \simeq 1.25 x^{\frac{1}{3}} \mbox{e}^{-x} \left(648 + x^2\right)^{\frac{1}{12}} \; .
\end{equation}
The average power of inverse-Compton Scattering (ICS) is given by
\begin{equation}
P_{IC} (\nu,E,z) = c E_{\gamma}(z) \int d\epsilon \; n(\epsilon) \sigma(E,\epsilon,E_{\gamma}(z)) \; ,
\label{eq:ics_power}
\end{equation}
where $E_{\gamma}(z) = h \nu (1+z)$ is the emitted photon energy, $n(\epsilon)$ is the black-body spectrum of the CMB photons, and $E$ is the electron energy. Here we consider mainly the ICS on CMB photons because this is the largest radiation background available in the universe.
Additionally,
\begin{equation}
\sigma(E,\epsilon,E_{\gamma}) = \frac{3\sigma_T}{4\epsilon\gamma^2}G(q,\Gamma_e) \; ,
\end{equation}
where $\sigma_T$ is the Thompson cross-section, $\gamma$ is the electron Lorentz factor, and
\begin{equation}
G(q,\Gamma_e) = 2 q \ln{q} + (1+2 q)(1-q) + \frac{(\Gamma_e q)^2(1-q)}{2(1+\Gamma_e q)} \; ,
\end{equation}
with
\begin{equation}
\begin{aligned}
q & = \frac{E_{\gamma}}{\Gamma_e(\gamma m_e c^2 + E_{\gamma})} \; , \\
\Gamma_e & = \frac{4\epsilon\gamma}{m_e c^2}
\end{aligned}
\end{equation}

Bremsstrahlung emission of DM-produced secondary electrons in a background atmosphere (usually the IGM, the ICM or the ISM)  has an average power
\begin{equation}
P_B (E_{\gamma},E,r) = c E_{\gamma}(z)\sum\limits_{j} n_j(r) \sigma_B (E_{\gamma},E) \; ,
\end{equation}
where $n_j(r)$ is the density of the background atmosphere species $j$, and
\begin{equation}
\sigma_B (E_{\gamma},E) = \frac{3\alpha \sigma_T}{8\pi E_{\gamma}}\left[ \left(1+\left(1-\frac{E_{\gamma}}{E}\right)^2\right)\phi_1 - \frac{2}{3}\left(1-\frac{E_{\gamma}}{E}\right)\phi_2 \right] \; ,
\end{equation}
with $\phi_1$ and $\phi_2$ being energy dependent factors determined by the species $j$(see \citep{longair1994}).

For the DM-induced $\gamma$-ray production the flux calculation is somewhat simplified
\begin{equation}
S_{\gamma} (\nu,z) = \int_0^r d^3r^{\prime} \, \frac{Q_{\gamma}(\nu,z,r)}{4\pi D_L^2} \; ,
\end{equation}
with $Q_{\gamma}(\nu,z,r)$ being the source function within the given DM halo.

The local emissivity for the $i-th$ emission mechanism  (synchrotron, ICS, bremsstrahlung) can then be found as a function of the electron and positron equilibrium distributions as well as the associated power
\begin{equation}
j_{i} (\nu,r,z) = \int_{m_e}^{M_\chi} dE \, \left(\td{n_{e^-}}{E}{} + \td{n_{e^+}}{E}{}\right) P_{i} (\nu,E,r,z) \; ,
\label{eq:emm}
\end{equation}
where $\td{n_{e^-}}{E}{}$ is the equilibrium electron distribution from DM annihilation/decay (see below).
The flux density spectrum within a radius $r$ is then written as
\begin{equation}
S_{i} (\nu,z) = \int_0^r d^3r^{\prime} \, \frac{j_{i}(\nu,r^{\prime},z)}{4 \pi D_L^2} \; ,
\label{eq:flux}
\end{equation}
where $D_L$ is the luminosity distance to the halo.

In electron-dependent emissions there are two processes of importance, namely energy-loss and diffusion. Diffusion is typically only significant within small structures~\citep{Colafrancesco2007,gsp2015}, thus must be accounted for within the environment of a UCMH.\\
The equilibrium electron distribution is found as a stationary solution to the equation
\begin{equation}
\begin{aligned}
\pd{}{t}{}\td{n_e}{E}{} = & \; \gv{\nabla} \left( D(E,\v{r})\gv{\nabla}\td{n_e}{E}{}\right) + \pd{}{E}{}\left( b(E,\v{r}) \td{n_e}{E}{}\right) + Q_e(E,\v{r}) \; ,
\end{aligned}
\end{equation}
where $D(E,\v{r})$ is the diffusion coefficient, $b(E,\v{r})$ is the energy loss function, and $Q_e(E,\v{r})$ is the electron source function from DM annihilation/decay. In this case, we will work under the simplifying assumption that $D$ and $b$ lack a spatial dependence and thus we will include only average values for magnetic field and thermal electron densities. For details of the solution see \citep{Colafrancesco2007}.\\
We thus define the functions as follows~\citep{Colafrancesco1998}
\begin{equation}
D(E) = \frac{1}{3}c r_L (E) \frac{\overline{B}^2}{\int^{\infty}_{k_L} dk P(k)} \; ,
\end{equation}
where $\overline{B}$ is the average magnetic field, $r_L$ is the Larmour radius of a relativistic particle with energy $E$ and charge $e$ and $k_L = \frac{1}{r_L}$. This combined with the requirement that
\begin{equation}
\int^{\infty}_{k_0} dk P(k) = \overline{B}^2 \; ,
\end{equation}
where $k_0 = \frac{1}{d_0}$, with $d_0$ being the smallest scale on which the magnetic field is homogeneous, yields the final form
\begin{equation}
D(E) = D_0 d_0^{\frac{2}{3}} \left(\frac{\overline{B}}{1 \mu\mbox{G}}\right)^{-\frac{1}{3}} \left(\frac{E}{1 \mbox{GeV}}\right)^{\frac{1}{3}}  \; , \label{eq:diff}
\end{equation}
where $D_0 = 3.1\times 10^{28}$ cm$^2$ s$^{-1}$, and we assume that $d_0 = 1 pc$ for the UCMHs we consider.

The energy loss function is defined by
\begin{equation}
\begin{aligned}
b(E) = & b_{IC} E^2 (1+z)^4 + b_{sync} E^2 \overline{B}^2 \\&\; + b_{Coul} \overline{n} (1+z)^3 \left(1 + \frac{1}{75}\log\left(\frac{\gamma}{\overline{n} (1+z)^3}\right)\right) \\&\; + b_{brem} \overline{n} (1+z)^3 \left( \log\left(\frac{\gamma}{\overline{n} (1+z)^3 }\right) + 0.36 \right) \;,
\end{aligned}
\label{eq:loss}
\end{equation}
where $\overline{n}$ is the average thermal electron density in the UCMH and is given in cm$^{-3}$, while $b_{IC}$, $b_{synch}$, $b_{col}$, and $b_{brem}$ are the inverse Compton, synchrotron, Coulomb and bremsstrahlung energy loss factors, taken to be $0.25$, $0.0254$, $6.13$, and $1.51$ respectively in units of $10^{-16}$ GeV s$^{-1}$. Here $E$ is the energy in GeV and the B-field is in $\mu$G. We will assume $ \overline{B} = 0.1$ $\mu$G, and $\overline{n} = 10^{-6}$ cm$^{-3}$ for the UCMHs we consider. These values are chosen in keeping with the low electron density of dwarf galaxies, and we choose the magnetic field an order of magnitude smaller than that typically used within a dwarf, making this a reasonable and slightly conservative basis on which to estimate synchrotron fluxes from DM processes.

\section{Detectability of UCMHs}
\label{sec:detect}

\begin{figure}[htbp]
\centering
\resizebox{0.6\hsize}{!}{\includegraphics[scale=0.7]{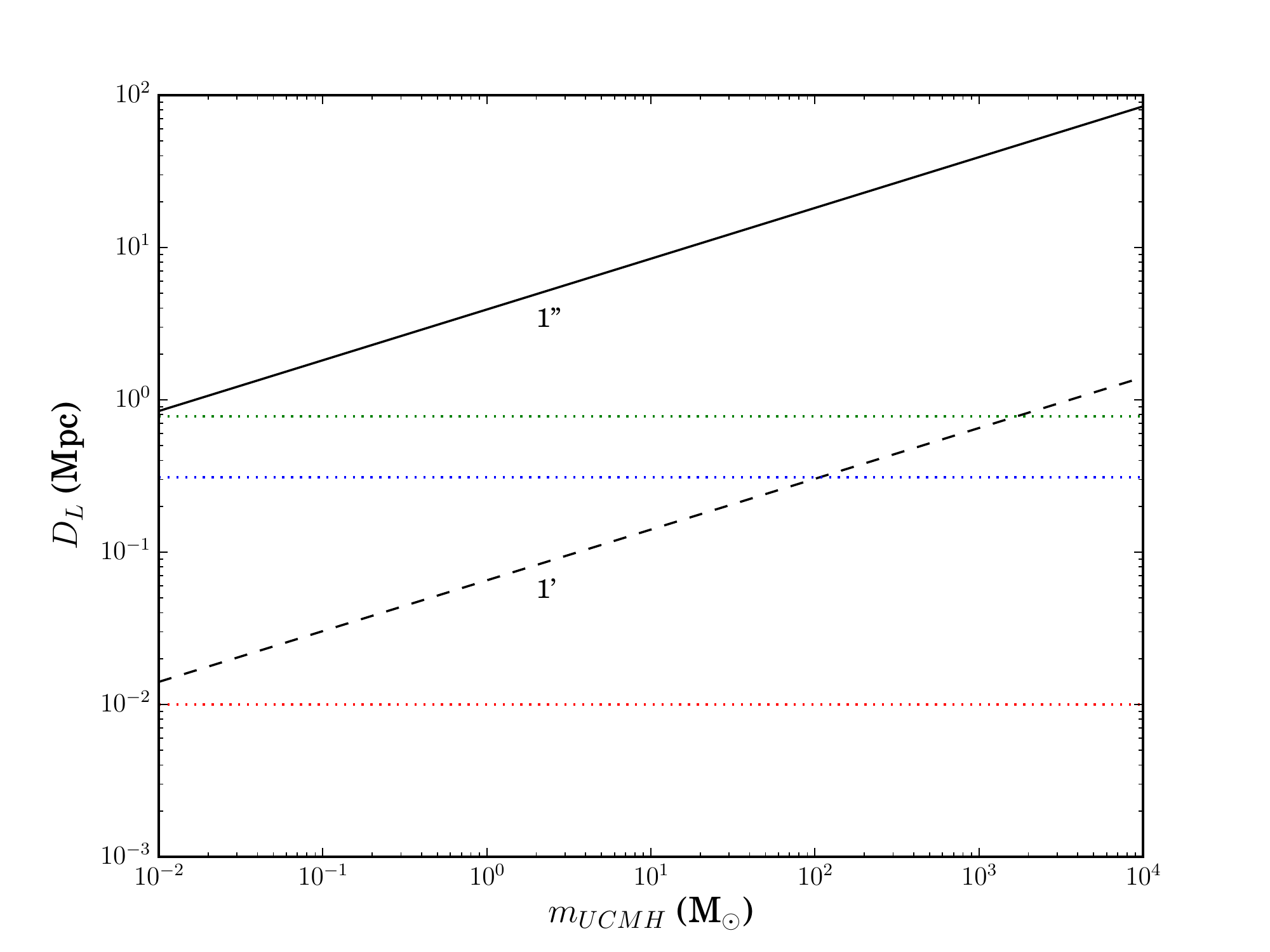}}
\caption{The distances at which UCMHs of various mass appear at arcminute and arcsecond angular sizes. The blue dotted line shows a distance of $310$ kpc from the Earth being taken as the Milky-Way virial radius~\citep{taylor2016}. The green dotted line shows the distance to Andromeda of $778$ kpc~\citep{distm31}, and the red dotted line shows the microlensing motivated 10 kpc observation radius.}
\label{fig:resolve}
\end{figure}

In this section we explore the detectability of UCMHs mainly in both radio and $\gamma$-ray frequency bands. This exploration will be confined to the region suggested for accurate microlens mass reconstruction by Gaia~\citep{gaia2001}, this being taken to be $\sim 10$ kpc. The first point to consider is whether such small structures can be resolved as point-like sources within these distances: to this aim we present Fig.~\ref{fig:resolve} to demonstrate that UCMHs of a wide variety of masses can be spatially resolved at the arcsecond (for radio) and arcminute level within the radius of the Milky-Way, and more easily within short observation distances that are viable for microlensing strategies.\\

\begin{figure}[htbp]
\centering
\resizebox{0.6\hsize}{!}{\includegraphics[scale=0.7]{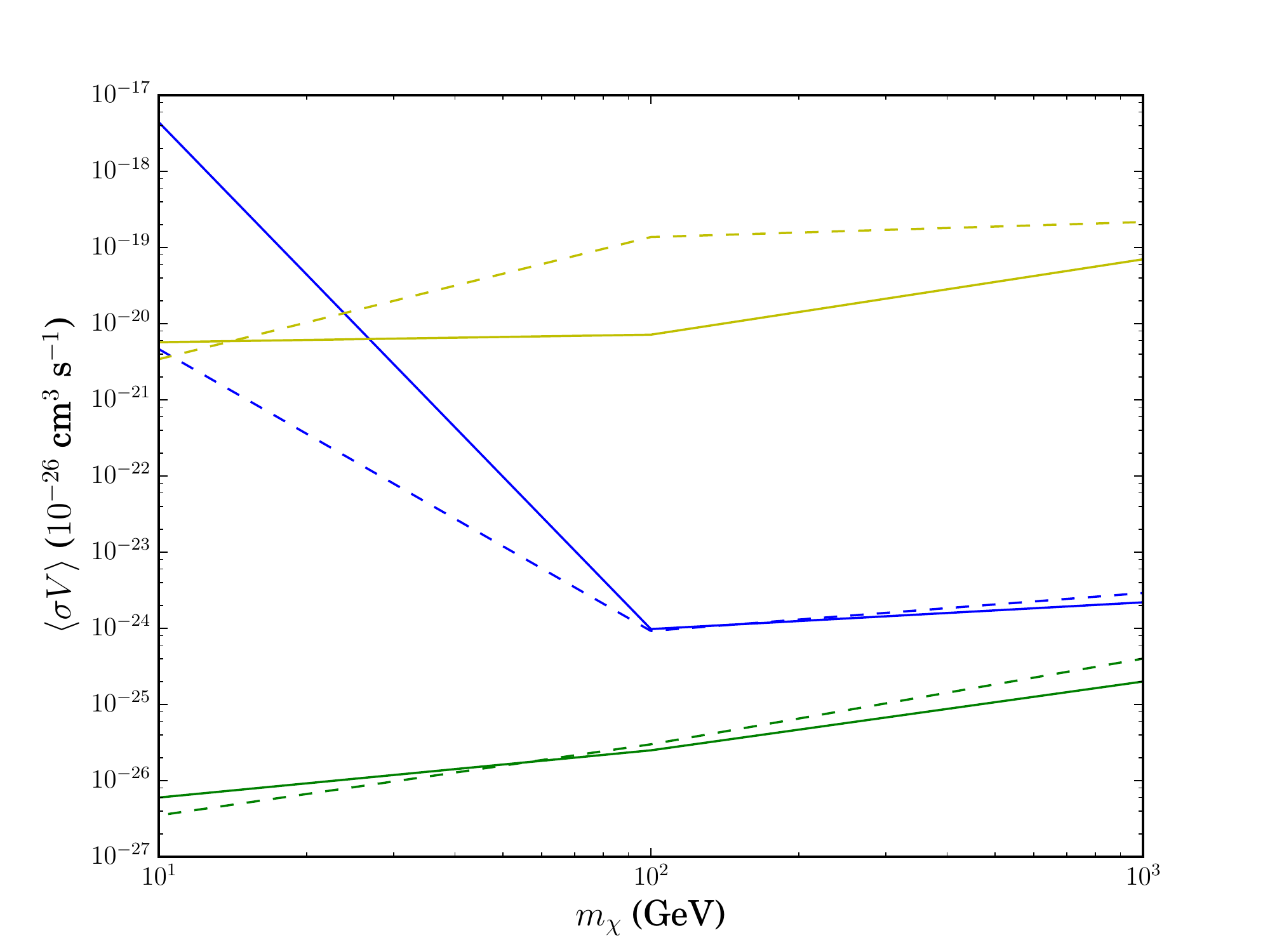}}
\caption{The cross-sections at which a UCMH of $10$ M$_{\odot}$ mass can be detected with a mJy flux by the SKA (blue) and nJy level by Fermi-LAT (yellow) in the $\tau^+\tau^-$ (dashed) and $b\bar{b}$ (solid) annihilation channels. The green lines show the current Fermi-LAT constraints from non-observation of $\gamma$-rays in dwarf galaxies~\citep{Fermidwarves2015}. The distance used is the 10 kpc radius motivated by Gaia.}
\label{fig:mjy1}
\end{figure}

\begin{figure}[htbp]
\centering
\resizebox{0.6\hsize}{!}{\includegraphics[scale=0.7]{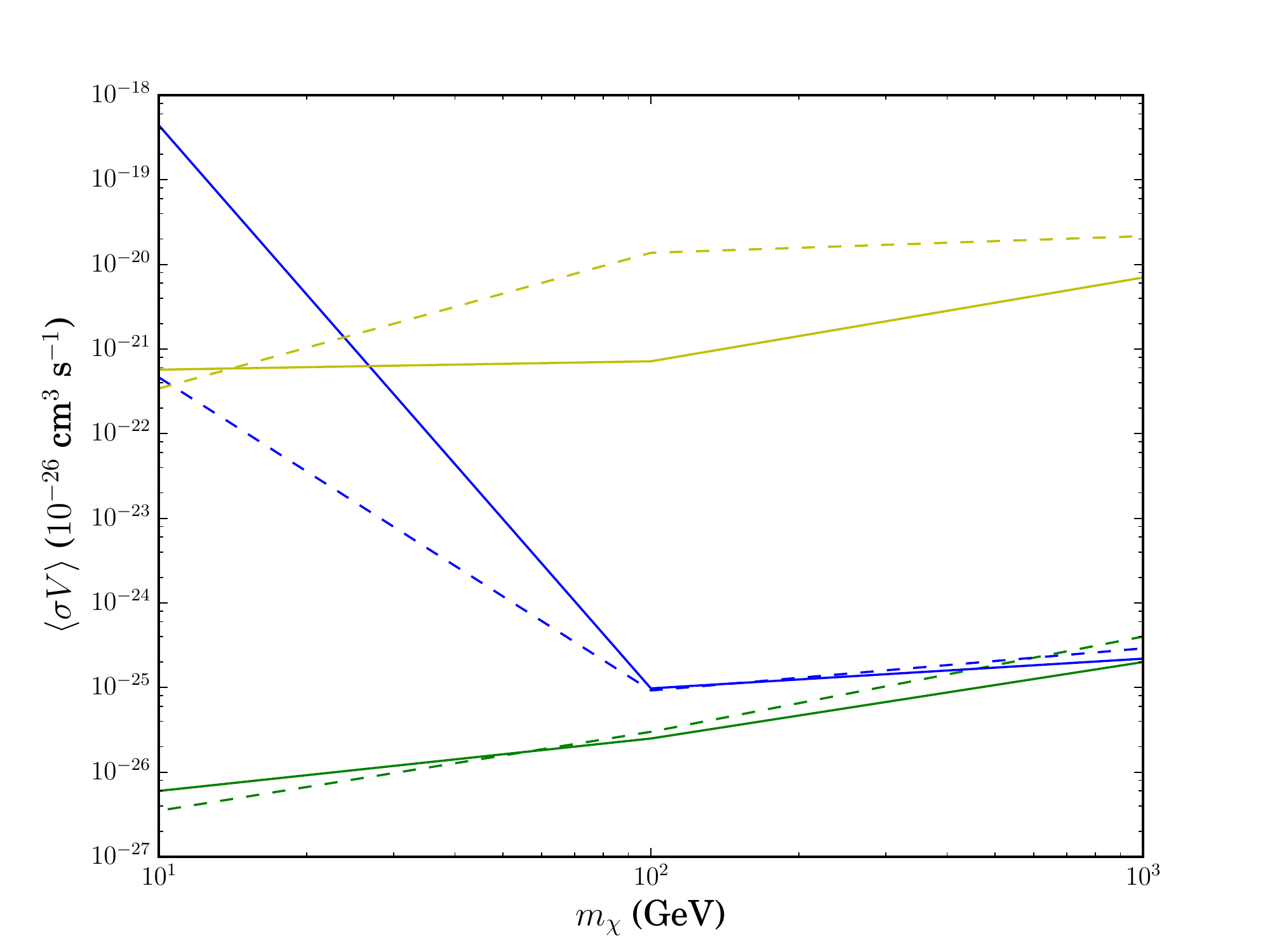}}
\caption{The cross-sections at which a UCMH of $10^2$ M$_{\odot}$ mass can be detected with a mJy flux by the SKA (blue) and nJy level by Fermi-LAT (yellow) in the $\tau^+\tau^-$ (dashed) and $b\bar{b}$ (solid) annihilation channels. The green lines show the current Fermi-LAT constraints from non-observation of $\gamma$-rays in dwarf galaxies~\citep{Fermidwarves2015}. The distance used is the 10 kpc radius motivated by Gaia.}
\label{fig:mjy2}
\end{figure}

\begin{figure}[htbp]
\centering
\resizebox{0.6\hsize}{!}{\includegraphics[scale=0.7]{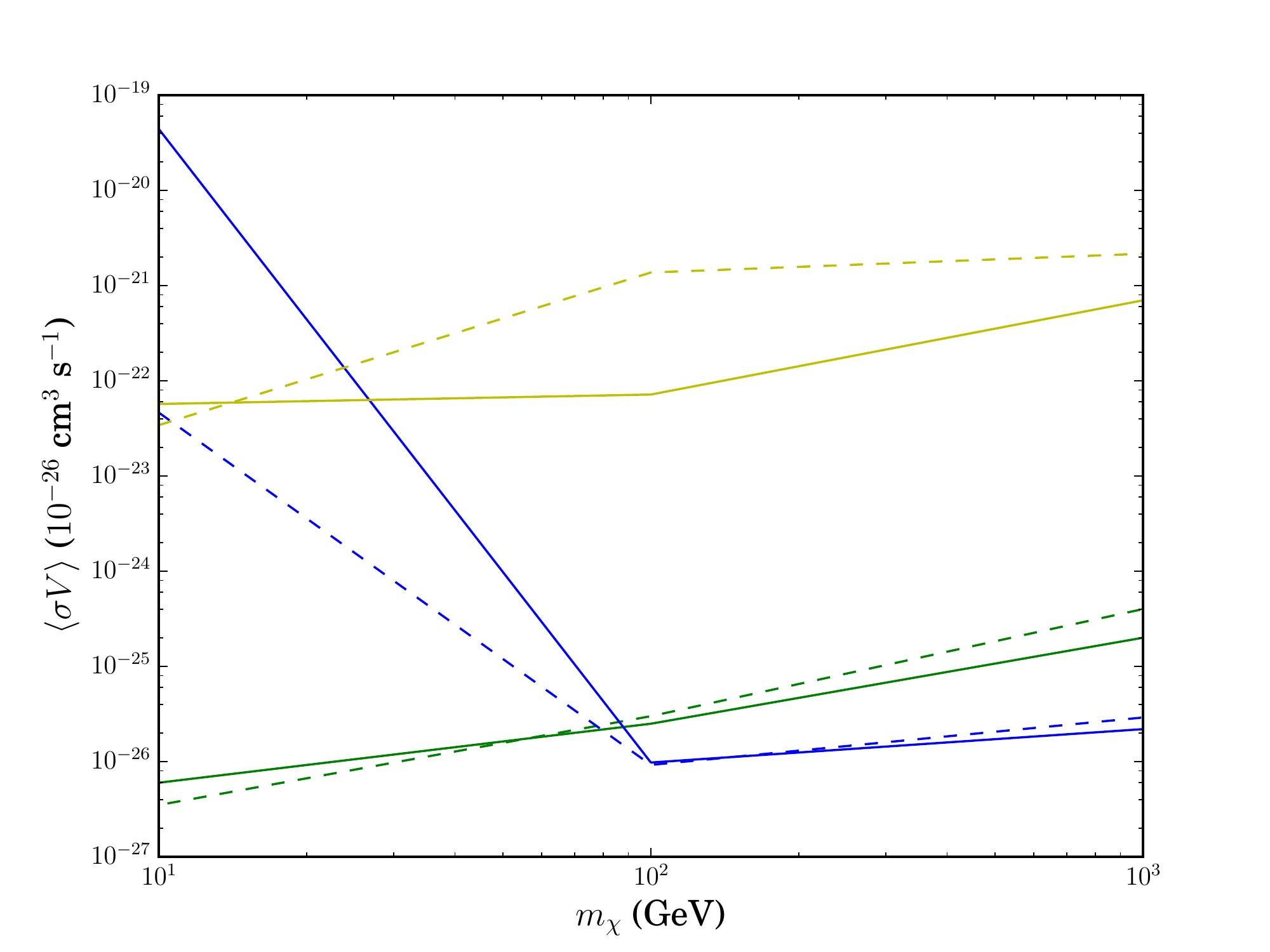}}
\caption{The cross-sections at which a UCMH of $10^3$ M$_{\odot}$ mass can be detected with a mJy flux by the SKA (blue) and nJy level by Fermi-LAT (yellow) in the $\tau^+\tau^-$ (dashed) and $b\bar{b}$ (solid) annihilation channels. The green lines show the current Fermi-LAT constraints from non-observation of $\gamma$-rays in dwarf galaxies~\citep{Fermidwarves2015}. The distance used is the 10 kpc radius motivated by Gaia.}
\label{fig:mjy3}
\end{figure}

The next question is which values of the annihilation cross-section (or decay rate) are necessary for DM emissions to be observable for UCMHs of various masses? In the case of annihilation, figures~\ref{fig:mjy1}, \ref{fig:mjy2}, and \ref{fig:mjy3} show the required cross-sections to observe both radio and $\gamma$-ray emissions at a mJy (for radio) and nJy (for $\gamma$-rays) at $5\sigma$ confidence levels as a function of the WIMP mass within the radius motivated by Gaia microlensing measures. The three figures correspond to a differing choice of UCMH mass, being $10$, $10^2$, and $10^3$ M$_{\odot}$, respectively. These observability requirements are compared to the existing cross-section limits placed by the Fermi-LAT study of dwarf galaxies~\citep{Fermidwarves2015}: this is the best significant choice of comparison limit given that Fermi-LAT has officially observed no UCMHs so far, so it is then consistent to infer that there are no UCMH candidates observed at the Fermi-LAT cross-section upper bounds (we note the Fermi-LAT null result on UCMHs is within a distance slightly greater than or equal to the region probed accurately by Gaia with microlensing). From these three figures we see that, for UCMH masses above $10^2$ M$_{\odot}$, radio observations are able to detect UCMHs at the mJy flux level with cross-sections below  Fermi-LAT dwarf limits; this is of particular significance as UCMHs seeded during the electron-positron annihilation epoch would have current epoch masses around $158$ \msol~\citep{josan2010}, and can be therefore proven with the SKA within the region that Gaia could search for candidates.\\
We note that none of these UCMHs are visible in $\gamma$-rays at the nJy level for any cross-section below or comparable to the Fermi-Lat dwarf limits, whether or not Fermi-LAT can prove the existence of these UCMHs will be considered below. A distinctive feature of the radio emissions is that low mass WIMPs of a few 10s of GeVs produce much lower synchrotron fluxes. This is because the $\sigv$, used to calculate the halo profile in Eq.~\ref{eq:rhomax}, was chosen independent of WIMP mass, but both of these contribute to define $\rho_{max}$. Although this would seem to advantage larger WIMP masses, in the case of the largest mass studied here (10 TeV) the variation caused by adjusting the cross-section choice in Eq.~\ref{eq:rhomax} by 2 orders of magnitude causes a flux drop by less than a factor of 3. Thus, the limits we derive are quite insensitive to our use of independent $\sigv$ and WIMP mass for the calculation of the halo profile (we stress that we apply mass dependent Fermi-LAT constraints on $\langle \sigma V \rangle$ to derive the final flux), and the inability of the SKA to constrain UCMH abundances for lower mass WIMPs, within the region searched by our dual methodology, is entirely due to the relative positions of the DM-induced synchrotron peak and the SKA's own sensitivity bands.\\

In the case of decaying WIMPs, none of the displayed UCMH masses is sufficient to be detectable at $\mu$Jy level in radio or nJy level in $\gamma$-rays with decay rates equal to or less than existing limits derived from Fermi-LAT extra-galactic background data~\citep{fermi-eg}.

\begin{figure}[htbp]
\centering
\resizebox{0.6\hsize}{!}{\includegraphics[scale=0.7]{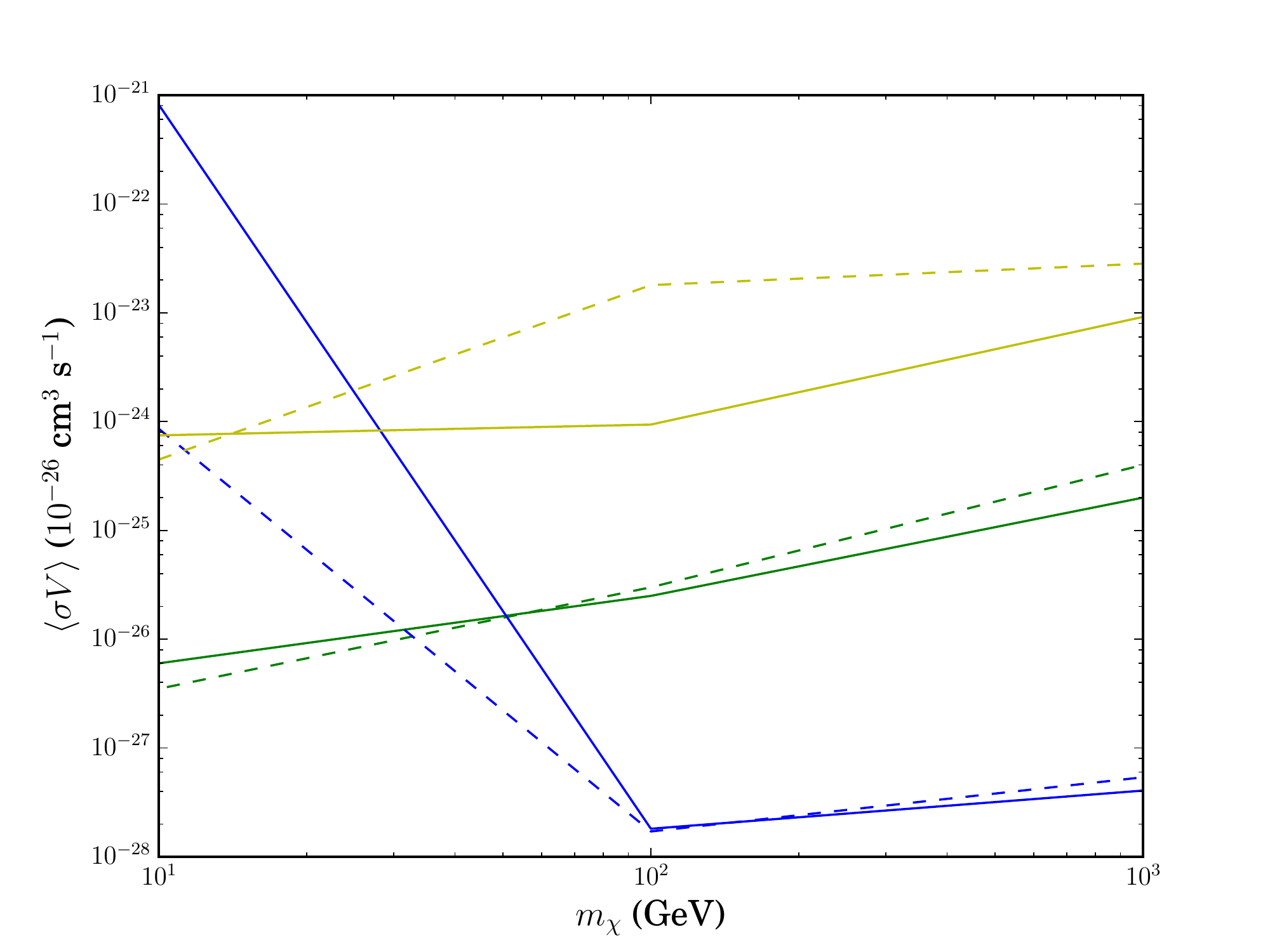}}
\caption{The cross-sections at which a UCMH of $10$ M$_{\odot}$ mass can be observed at the minimum detectable flux by the SKA (blue) and Fermi-LAT (yellow) in the $\tau^+\tau^-$ (dashed) and $b\bar{b}$ (solid) annihilation channels. The green lines show the current Fermi-LAT constraints from non-observation of $\gamma$-rays in dwarf galaxies~\citep{Fermidwarves2015}. The distance used is the 10 kpc radius motivated by Gaia.}
\label{fig:max1}
\end{figure}

\begin{figure}[htbp]
\centering
\resizebox{0.6\hsize}{!}{\includegraphics[scale=0.7]{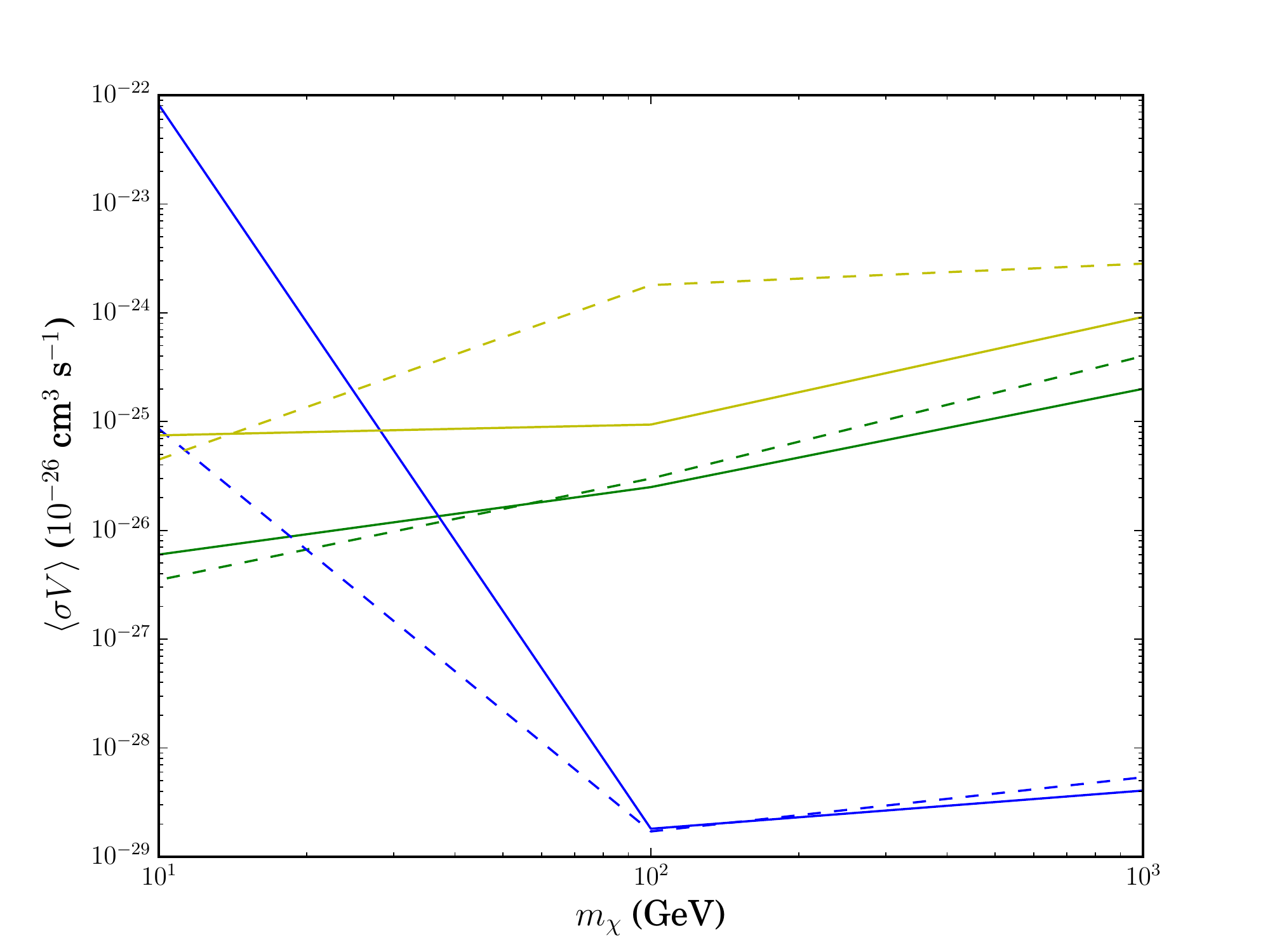}}
\caption{The cross-sections at which a UCMH of $10^2$ M$_{\odot}$ mass can be observed at the minimum detectable flux by the SKA (blue) and Fermi-LAT (yellow) in the $\tau^+\tau^-$ (dashed) and $b\bar{b}$ (solid) annihilation channels. The green lines show the current Fermi-LAT constraints from non-observation of $\gamma$-rays in dwarf galaxies~\citep{Fermidwarves2015}. The distance used is the 10 kpc radius motivated by Gaia.}
\label{fig:max2}
\end{figure}

\begin{figure}[htbp]
\centering
\resizebox{0.6\hsize}{!}{\includegraphics[scale=0.7]{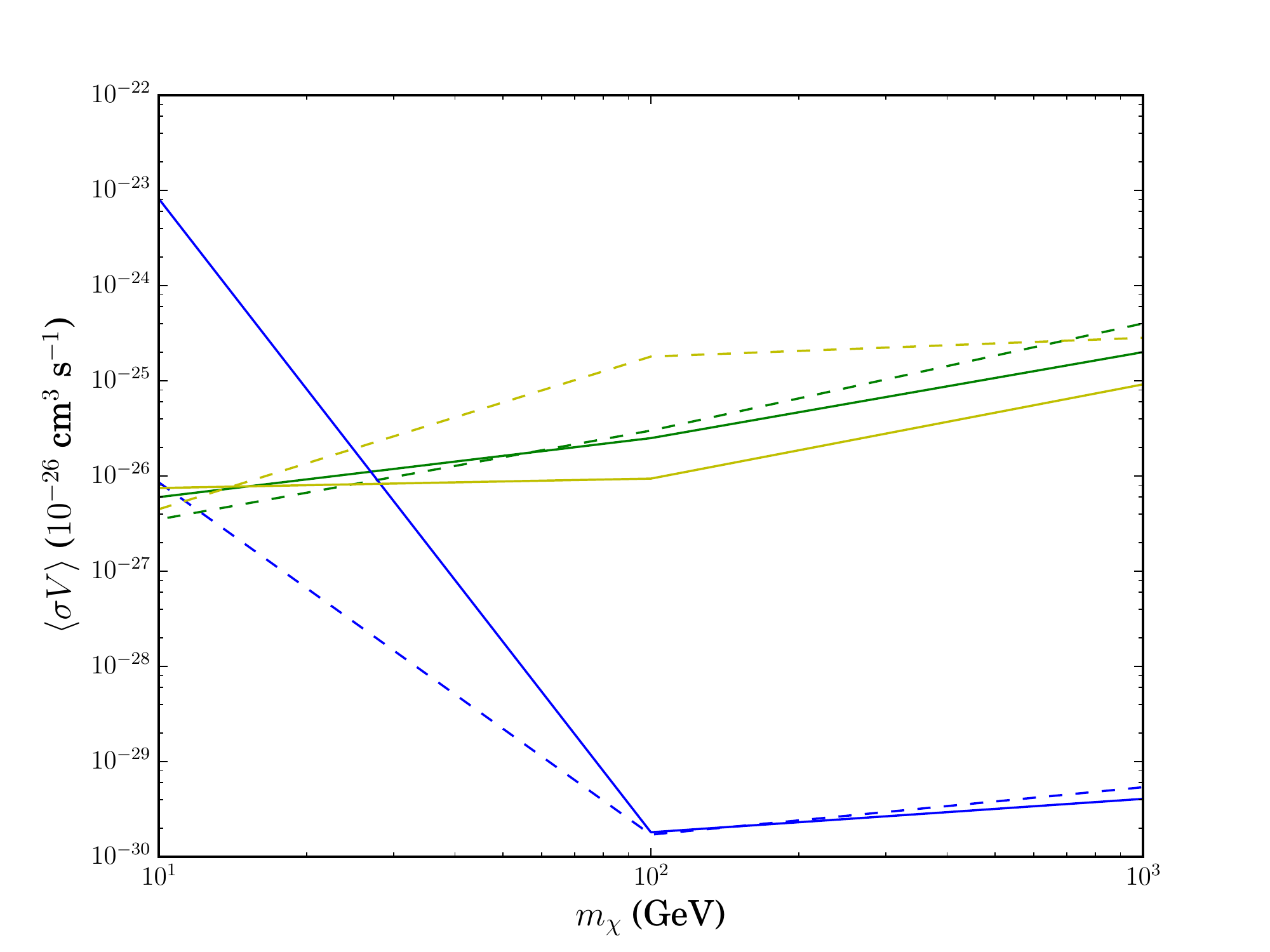}}
\caption{The cross-sections at which a UCMH of $10^3$ M$_{\odot}$ mass can be observed at the minimum detectable flux by the SKA (blue) and Fermi-LAT (yellow) in the $\tau^+\tau^-$ (dashed) and $b\bar{b}$ (solid) annihilation channels. The green lines show the current Fermi-LAT constraints from non-observation of $\gamma$-rays in dwarf galaxies~\citep{Fermidwarves2015}. The distance used is the 10 kpc radius motivated by Gaia.}
\label{fig:max3}
\end{figure}

To see how far this can be pushed, we again determine the same cross-section (or decay rate) requirements, but this time requiring minimal fluxes at the $5\sigma$ detection limits of the SKA with 1000 hours of observing time and Fermi-LAT Pass 8 sensitivities. These are displayed for $10$, $10^2$, and $10^3$ M$_{\odot}$ UCMHs in figs.~\ref{fig:max1}, \ref{fig:max2}, and \ref{fig:max3}, respectively. For all three UCMH masses, the SKA can observe them in radio for WIMP masses above a few 10s of GeVs. The minimal detectable cross sections reach as low as  $10^{-28}$ cm$^3$ s$^{-1}$ for WIMPs between 100 GeV and 1 TeV in the case of the UCMHs above $10^2$ \msol. With its most sensitive configuration, Fermi-LAT (using the Pass 8 10yr differential sensitivity) shows hints of being able to detect UCMHs above a mass of $10^2$ \msol at cross-sections below the existing dwarf galaxy limits. This provides some suggestion towards why the Fermi-LAT has not officially observed any UCMH candidates, as the detection possibility looks to be rather marginal for $M_{UCMH} \leq 10^2 M_{\odot}$ even for the comparatively small 10 kpc search region used here. This provides some minor support to the arguments, based on the identification of promising unidentified Fermi-LAT point sources with Blazars, put forward in \citep{bringmann2012,zechlin2012} that none of the candidates suggested in various studies~\citep{buckley2010,belikov2012,zechlin2012} (as well as arXiv:1007.2644) are viable, but does not serve by itself to rule out these possibilities.

\begin{figure}[htbp]
\centering
\resizebox{0.6\hsize}{!}{\includegraphics[scale=0.7]{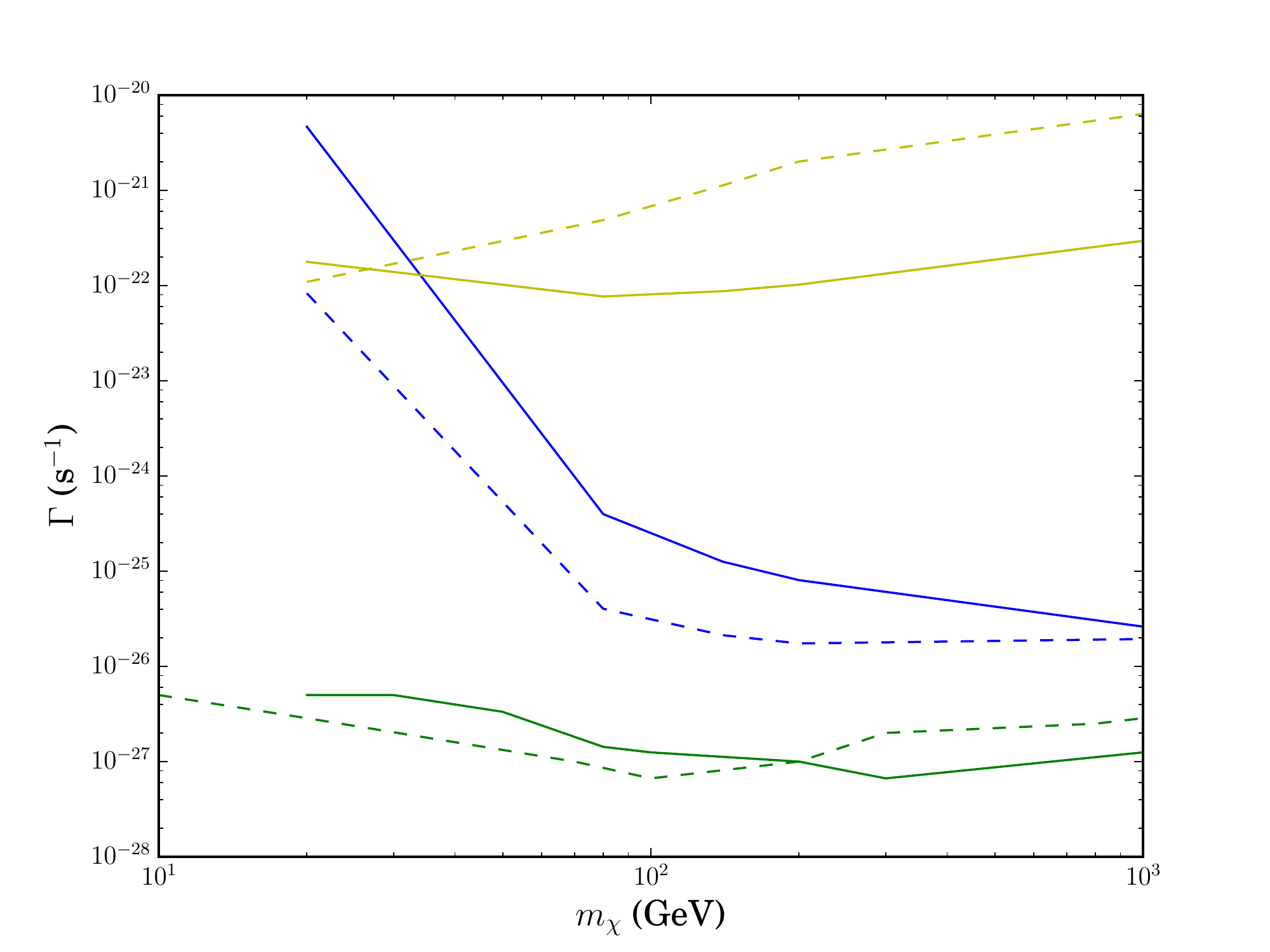}}
\caption{The decay rates at which a UCMH of $10^3$ M$_{\odot}$ mass can be observed at the minimum detectable flux by the SKA (blue) and Fermi-LAT (yellow) in the $\tau^+\tau^-$ (dashed) and $b\bar{b}$ (solid) decay channels. The green lines show the current Fermi-LAT constraints drawn from extra-galactic background data~\citep{fermi-eg}. The distance used is the 10 kpc radius motivated by Gaia.}
\label{fig:max_decay}
\end{figure}

In the case of decaying WIMPs, shown in fig.~\ref{fig:max_decay}, none of the displayed UCMH masses is sufficient to be reliably detectable at the maximal sensitivity levels of either the SKA or Fermi-LAT. In the best case scenario WIMPs around 1 TeV or more in mass would have visible decay emissions only from halos above $10^4$ solar masses. Once again this assumes decay rates equal to the upper bounds from Fermi-LAT extra-galactic background data~\citep{fermi-eg}.

\section{Constraints from non-observation in the Milky-Way}
\label{sec:large}
Having addressed the observability of individual UCMHs in radio and at very high energies, within a local region that can also be probed accurately for microlensing objects by Gaia. We move on to the interdependence of constraints on the maximum UCMH fraction $f_{max}$ and the cross-section $\langle \sigma V \rangle$. In particular we aim to examine how sensitive the limits on $f_{max}$ are to the sensitivity range of a given experiment and to a chosen value of $\sigv$, given that we have specified an observation radius motivated by external empirical considerations. This analysis is only performed for annihilating DM due to the weak detection prospects for decay in UCMHs shown above (see fig.~\ref{fig:max_decay}).

\begin{figure}[htbp]
\centering
\resizebox{0.6\hsize}{!}{\includegraphics[scale=0.7]{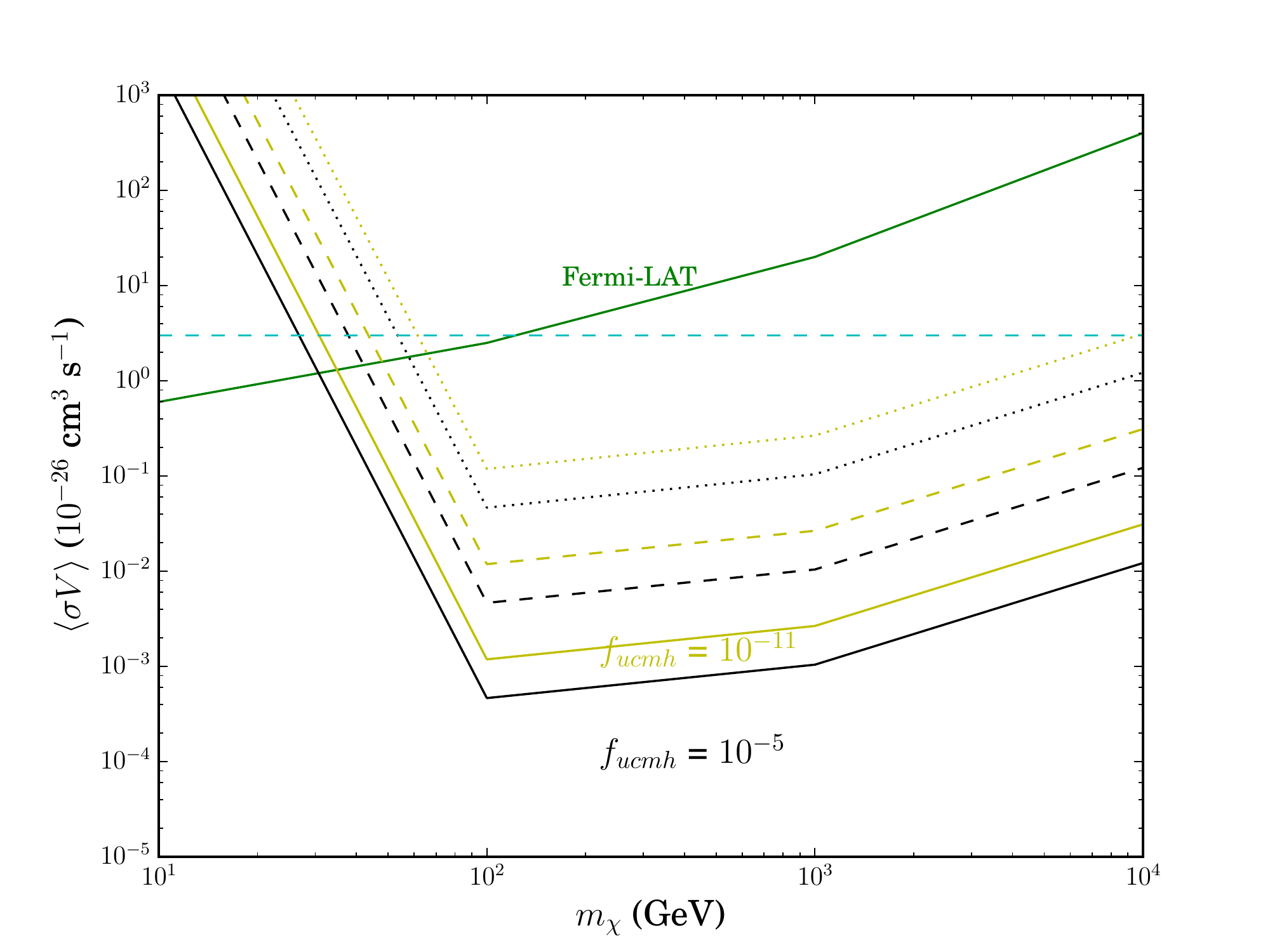}}
\caption{The cross-section constraints, on the $b\bar{b}$ channel, for a non-observation of a UCMH within 10 kpc by the SKA, given for a set of values of the UCMH abundance fraction. Solid lines show halo mass $10^3$ M$_{\odot}$, dashed $10^2$ M$_{\odot}$, and dotted $10$ M$_{\odot}$. The green line shows the Fermi-LAT constraint from dwarf galaxies and the cyan dashed line shows the thermal relic cross-section.}
\label{fig:sigv_fmax}
\end{figure}

In Figure~\ref{fig:sigv_fmax} we show the smallest values of the cross-section that can be probed in a UCMH search with the SKA, for a set of chosen $f_{UCMH}$ values, within the 10 kpc radius motivated by both indirect and microlensing considerations (we use the $b\bar{b}$ channel as a reference case). There are two dependencies here that affect how sensitive we are to different values of $\sigv$, these being the UCMH mass and abundance $f$. It is clear that not only is the dependence of $\sigv$ on the halo mass stronger than that on $f$, but that WIMP mass also has a more significant influence. Within the given radius, the SKA suffers from inability to probe WIMP models of a few tens of GeV and below at a level below the existing Fermi-LAT constraints from dwarf galaxies. This is due to the position of the DM-induced synchrotron radiation peak within the sensitivity band of the SKA. These results suggest that there is no strong degeneracy between the cross-section that can be probed and subsequent limits that are set on the UCMH abundance.

\begin{figure}[htbp]
\centering
\resizebox{0.6\hsize}{!}{\includegraphics[scale=0.7]{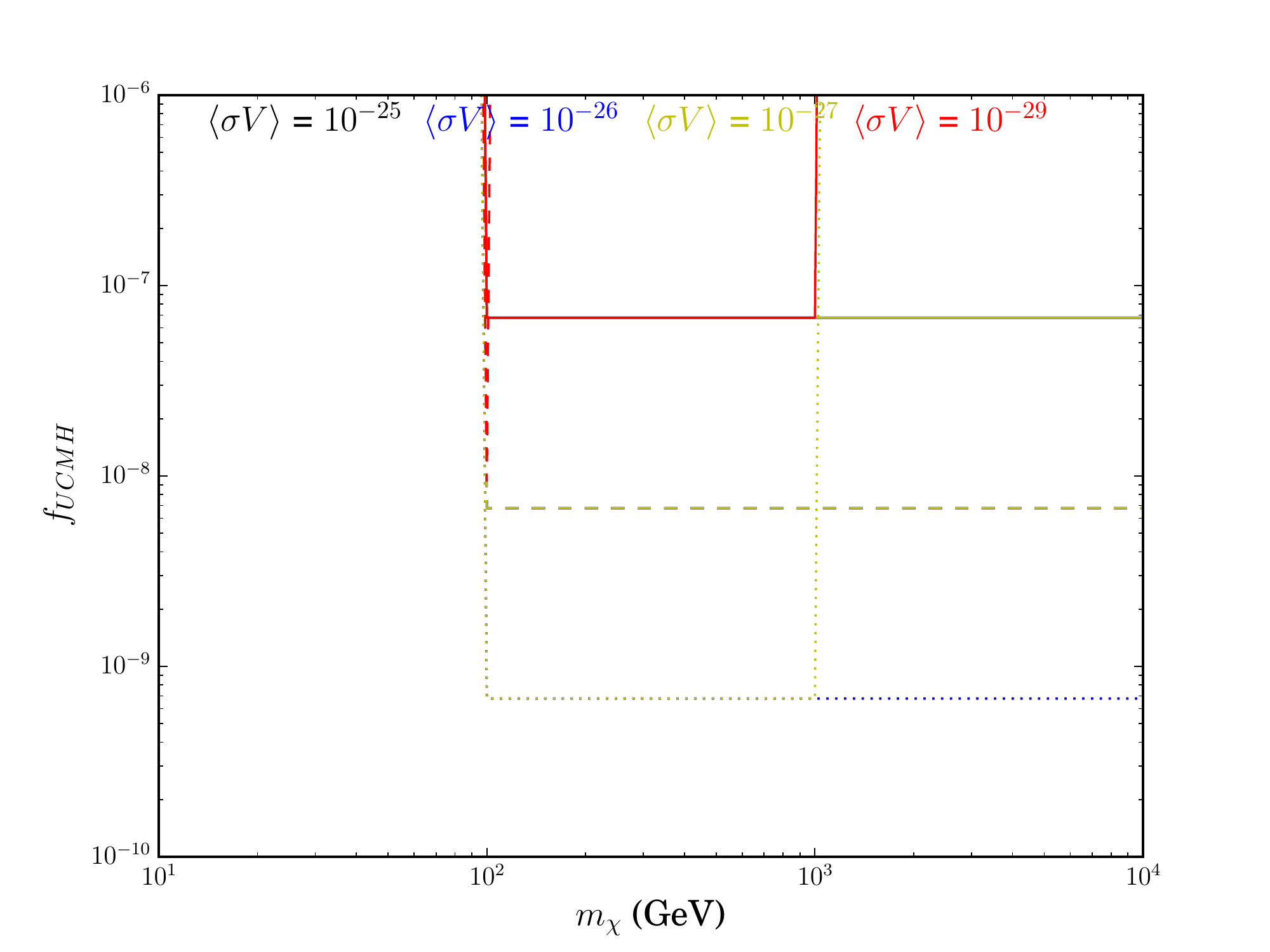}}
\caption{Constraints on the UCMH abundance given various values of the $b\bar{b}$ annihilation cross-section in the event of the SKA observing no UCMH candidates within 10 kpc.  Solid lines show halo mass $10^3$ M$_{\odot}$, dashed $10^2$ M$_{\odot}$, and dotted $10$ M$_{\odot}$. }
\label{fig:fmax_sigv}
\end{figure}

Of course, this exercise can also be done the other way around, as shown in Fig.~\ref{fig:fmax_sigv}. In this figure, we see that the dominant effects on UCMH detectability are the masses of the UCMH and of the WIMP, with $\sigv$ having smaller effect on the resulting limits. The effects of $\sigv$ can be seen in the fact that smaller cross-sections show a weakening in high WIMP mass constraints, the most severe seen for the $10^2$ \msol halos for the weakest displayed cross-section value. However, it is important to recognize that, for lower WIMP mass, the resulting abundance constraints are no different to those at other cross-sections. This serves to demonstrate that the SKA can probe the abundance of these compact targets well below the levels of Fermi-LAT upper limits on cross-sections. More importantly it demonstrates that the annihilation cross-section and UCMH abundance are not entirely degenerate. The asymmetry between the inter-dependencies of $\sigv$ and $f$ is entirely due to the nature of Eq.(\ref{eq:fmax}), as $\sigv$ enters only in the calculation of the probability $\sigma_{exp}$ (as $d_{obs}$ is chosen independently on physical grounds). The nature of the inter-dependence of $\sigv$ and $f$ is highly important, as if a large population of UCMHs can easily escape constraint through a small WIMP annihilation cross-section the limits placed on $f$ in this way are of little real impact. Thus, we have shown that, if $d_{obs}$ is chosen based on physical motivations independent of indirect detection sensitivities, it is possible to obviate the problematic degeneracy between $f$ and $\langle \sigma V\rangle$.

\begin{figure}[htbp]
\centering
\resizebox{0.6\hsize}{!}{\includegraphics[scale=0.7]{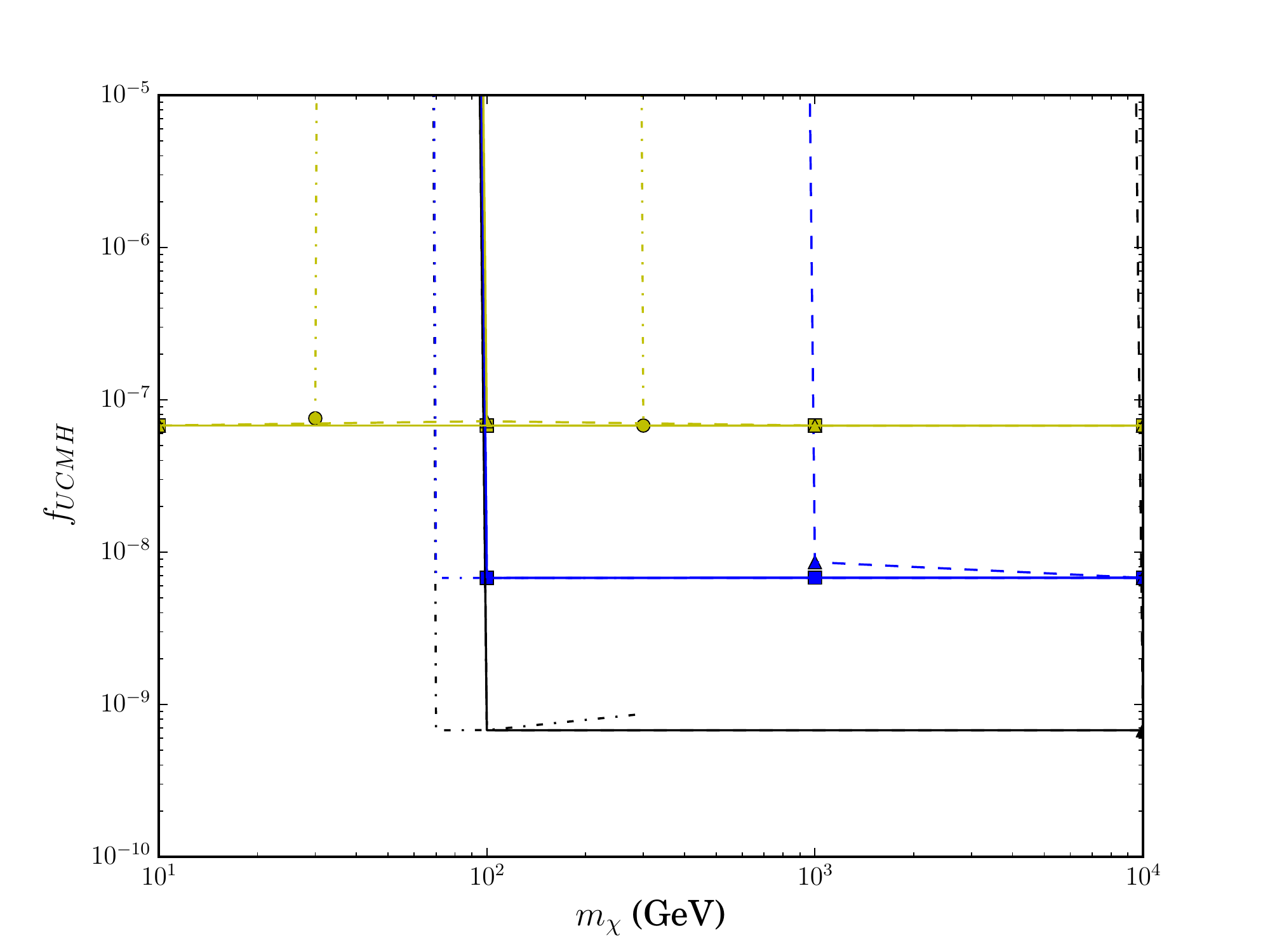}}
\caption{Constraints on UCMH abundance derived using the Fermi-LAT dwarf cross-section, given that Fermi has observed no UCMH candidates. In this case the SKA lines are shown just for a sensitivity comparison. SKA projections are given by the lines, with solid lines showing the $b\bar{b}$ channel, dashed show $\tau^+\tau^-$, and dash-dotted show the $hh$ case. Fermi-LAT is shown in points, with circles showing $hh$, squares $b\bar{b}$, and triangles for $\tau^+\tau^-$. The halo masses for both experiments are as follows: black (halo mass $10$ M$_{\odot}$), blue (halo mass $10^2$ M$_{\odot}$), and yellow (halo mass $10^3$ M$_{\odot}$). }
\label{fig:fucmh_fermi_ska}
\end{figure}

In Figure~\ref{fig:fucmh_fermi_ska} this analysis is performed with the cross-section fixed to the Fermi-LAT dwarf limits and then the consequences for $f_{max}$ of the SKA and Fermi observing no UCMHs are determined. In this figure we see that the details of experiment (and emission frequency) limit the abundance constraints by setting a minimal mass limit, below which no constraint can be derived. The SKA suffers substantially from this against Fermi-LAT for smaller WIMP masses, but the SKA can both perform equally to Fermi-LAT at higher masses and probe substantially smaller halos (Fermi-LAT cannot probe below UCMH masses of $10^2$ \msol for the given cross-sections in $b\bar{b}$ and $\tau^+\tau^-$ channels, while for $hh$ it cannot go below $10^3$ \msol). Particularly, the $b\bar{b}$ channel always displays the 100 GeV cut-off mass for smaller halos, while this is not so for $\tau^+\tau^-$ and $hh$, with the former only displaying a more halo-mass dependent cut-off, while the latter has a soft cut-off near 70 GeV for $< 10^2$ \msol halos and a halo-mass dependent upper limit on constrainable WIMP masses.

For decaying WIMPs no constraints can be placed on abundance when the maximal radius is set to 10 kpc, the decay fluxes are too small at this distance to be reliably detected.

To explore the future prospects for UCMH limits from SKA and CTA (in annihilation only), we then calculate the flux for the Reticulum II dwarf galaxy using  the J-factor from \citep{bonnivard2015} for high-energy flux and for radio fluxes. We choose Reticulum II in particular because it provides us with an independent cross-section limit from one of the highest J-factor dwarf galaxies yet discovered. Thus, we can expect some of the best future limits on the annihilation cross-section to be found with heavily DM dominated dwarf galaxy targets like Reticulum II. To model the DM distribution in Reticulum II we use an NFW profile normalized to the J-factor while assuming $\bar{B} = 1$ $\mu$G, $\overline{n} = 10^{-6}$ cm$^{-3}$, in keeping with values used for dwarf galaxies in the literature~\citep{Colafrancesco2006}. Then the null-constraints on $\langle \sigma V\rangle$ for CTA and SKA observations are determined at the $3\sigma$ level. Using these as the independent cross-section limits, we then construct the resulting values for $f_{max}$ assuming that neither experiment observes any UCMHs. The same exercise is performed for Fermi-LAT, but, we use the existing dwarf limits (which are slightly better than those directly set from Reticulum II itself). Finally, given that there are no other processes in the structure of the UCMH that will produce $\gamma$-rays, a promising strategy would be to use the CTA $\sigv$ limits from Reticulum II (as these contain the fewest uncertainties), with the high sensitivity of the SKA and microlensing probes used to search for UCMHs via correlation of radio point sources and unknown lensing objects.

\begin{figure}[htbp]
\centering
\resizebox{0.6\hsize}{!}{\includegraphics[scale=0.7]{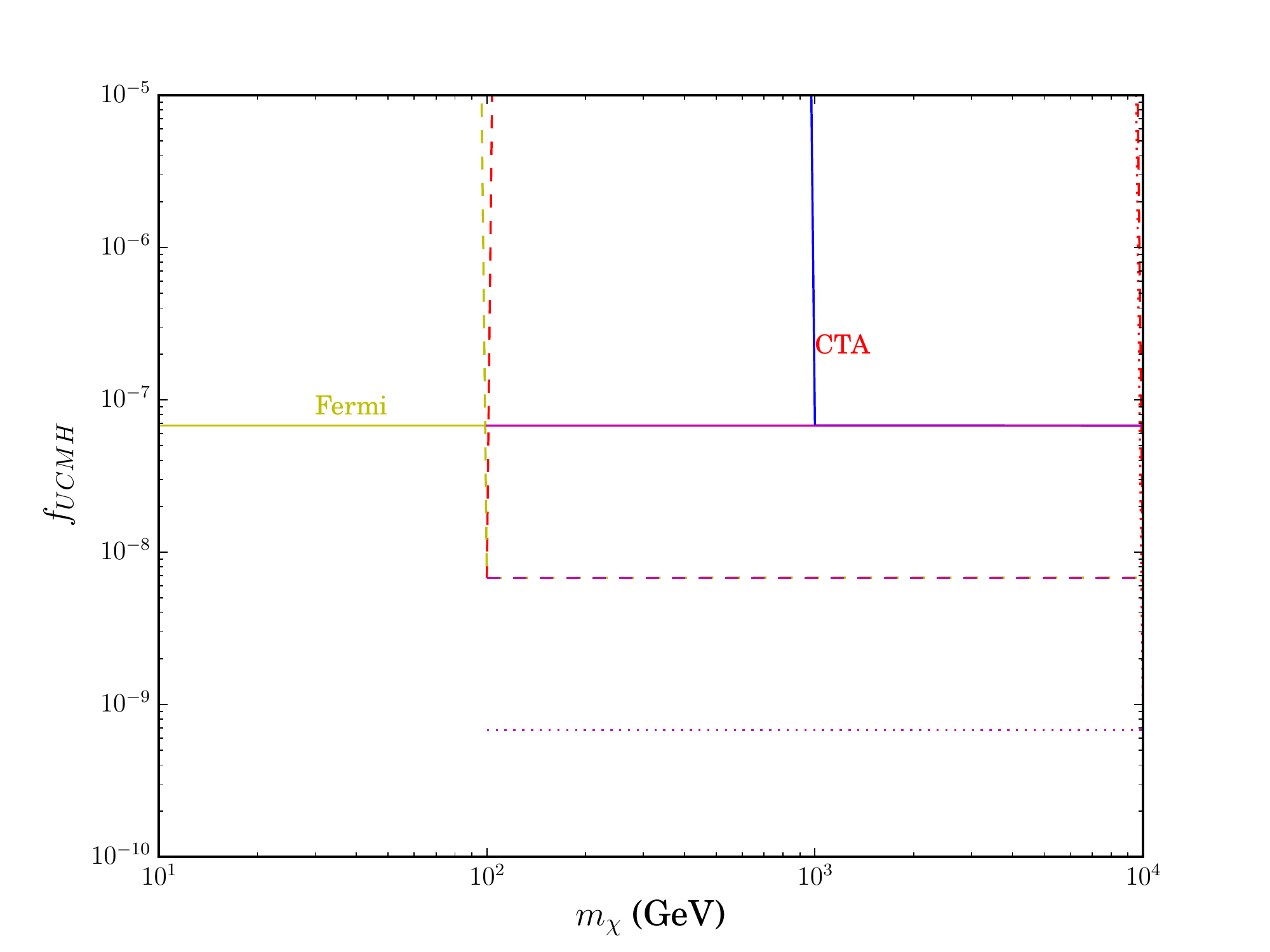}}
\caption{Non-detection constraints derived using $b\bar{b}$ cross-sections that assume a null result in the Reticulum 2 dwarf galaxy and that no UCMH candidates are detected by both the instrument and the Gaia mission. The SKA is shown in blue, Fermi in yellow, CTA in red. Solid lines show halo mass $10^3$ M$_{\odot}$, dashed $10^2$ M$_{\odot}$, and dotted $10$ M$_{\odot}$.}
\label{fig:fmax_bb}
\end{figure}

Fig.~\ref{fig:fmax_bb} demonstrates that the same pattern persists from previous figures. The SKA is limited in its ability to probe the lower mass WIMP models within UCMHs, but the abundance constraints are largely insensitive to the cross-section or experimental frequency window above this cut-off mass. CTA also shows an upper-bound on WIMP masses at 10 TeV for halos smaller than $10^3$ \msol. Despite the WIMP mass limitations, both CTA and SKA can probe abundances for smaller halo masses than Fermi-LAT. When the $hh$ and $\tau^+\tau^-$ channels are studied, the same pattern of variation between the channels seen in Fig.~\ref{fig:fucmh_fermi_ska} is replicated.

%

\section{Cosmological constraints}
\label{sec:cosmo}

Because UCMHs are primordial structures that can survive basically un-altered up to the present epoch, we can use their local abundance to set constraints on the primordial power-spectrum of the fluctuation field from where they originated. Due to the weak possibility of setting abundance limits for decaying WIMPs (see previous section) we will only focus on annihilating scenarios in this section. However, this analysis can be generically applied to any WIMP mass with a corresponding limit on $f$, regardless of decaying/annihilating phenomenology.

It is possible to place limits on allowed values of $\alpha_s$ and $\beta_s$ (from Eq.~(\ref{eq:pspec})) in a simple manner. First, we take $n_s$ to be well constrained by Planck~\citep{planck2014} to be $n_s (k_*) = 0.964$, and require that the tensor-to-scalar ratio is bound by $r_{ts} < 0.11$~\citep{planck2015}. Then for each pair of $\alpha_s$, $\beta_s$ we calculate $\beta_U$ for a given UCMH wave-number $k_R$ and compare the resulting UCMH fraction $f$ to the $3\sigma$ $f_{max}$ resulting from Fermi-LAT and SKA non-observation of a UCMH on the same scale; we  assume $m_{\chi} = 1$ TeV and  $f_{max}$ taken from the results in Fig.~\ref{fig:fucmh_fermi_ska}. If the running values $\alpha_s$ and $\beta_s$ result in $f > f_{max}$ then we take this pair to be excluded at $3\sigma$ confidence level.

\begin{figure}[htbp]
\centering
\resizebox{0.6\hsize}{!}{\includegraphics[scale=0.7]{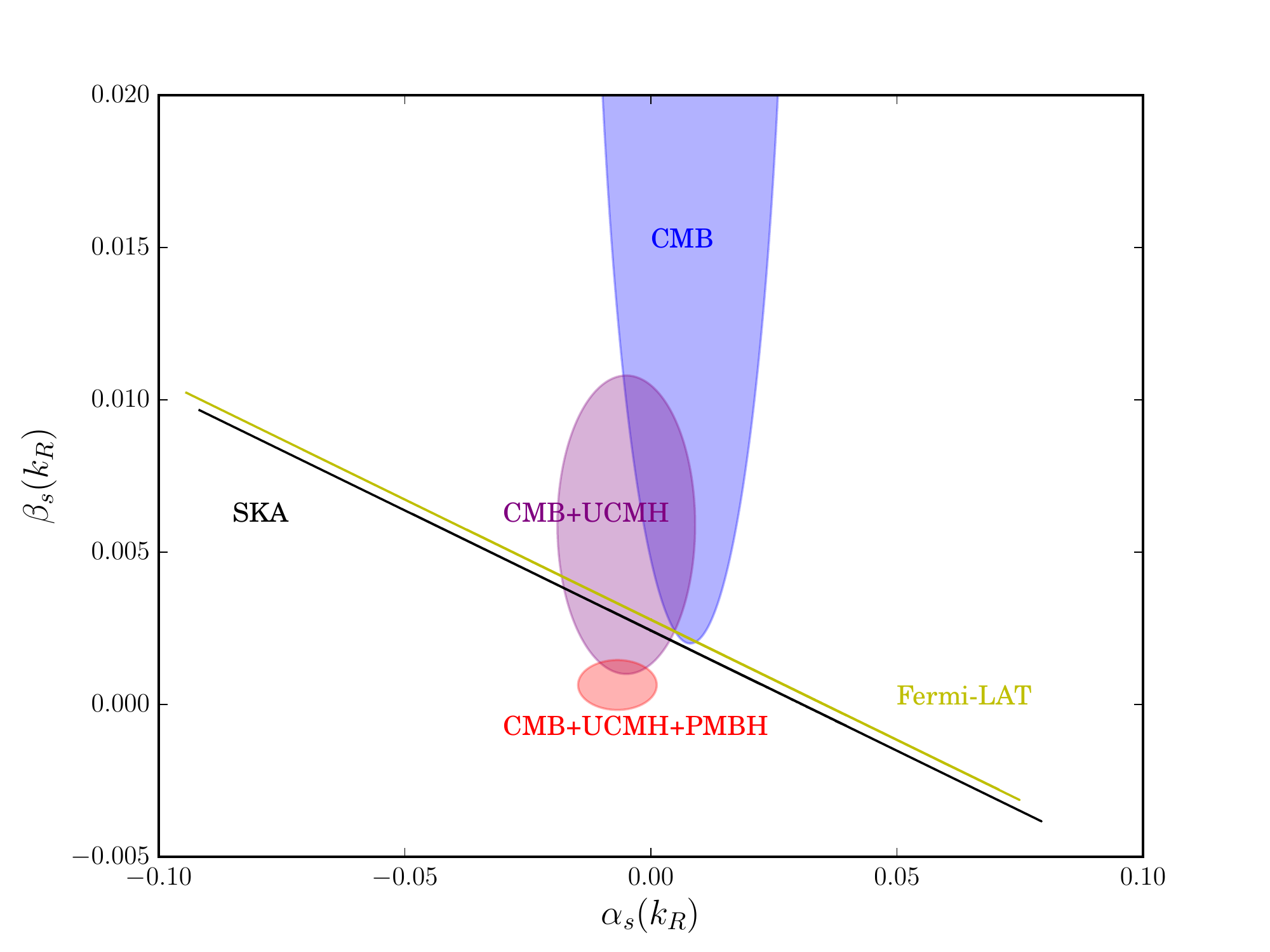}}
\caption{$3\sigma$ Upper limits at $k_R = 15874$ Mpc$^{-1}$ on $\alpha_s (k_R)$ and $\beta_s (k_R)$, the local running of power spectrum index $n_s$ and the running of the running respectively. The yellow and black lines show Fermi-LAT and SKA constraints respectively. The ellipses show the $3\sigma$ favoured regions from~\citep{aslanyan2015}, with plain Planck data in blue, Planck and Fermi-LAT UCMH constraints in purple, while the red additionally includes primordial black hole constraints.}
\label{fig:pspec}
\end{figure}

As an example we display the results for $k_R = 15874$ Mpc$^{-1}$ in Fig.~\ref{fig:pspec} compared to the $3\sigma$ favoured regions from \citep{aslanyan2015}. This demonstrates that we can provide strong constraints on the local running of the primordial power spectrum with this method, as we rule out the CMB only region and most of the CMB and UCMH favoured region. We note that UCMHs thus provide the ability to constrain the primordial power spectrum of perturbations on much smaller scales than those sampled by the CMB, which is limited to $k \sim 0.1$ Mpc$^{-1}$ at most. What is also evident is that the large difference in sensitivity between the SKA and Fermi-LAT is not reflected in the constraints on the primordial power spectrum, which only display a small improvement for the SKA case. This is because of the similarity between the limits derived by SKA and Fermi-LAT non-observation seen above (Fig.~\ref{fig:fucmh_fermi_ska}), as well as the fact that minor adjustments of $\alpha_s (k_R)$ and $\beta_s (k_R)$ can result in relatively large fluctuations of $f$ at the large values of $k$ used ($k \sim 10^{4}$ Mpc$^{-1}$) as a result of the dependencies of Eq.~(\ref{eq:delta-chi}).

\section{Local Solar system constraints}
\label{sec:local}
In this section we will cover the results for the effects of DM capture by Earth and Mars. In particular we will use possible changes in the core temperature, as well as relative increases in the rate of heat-flow out of the core, as metrics of possible increases in volcanism on the planet in question. This is in order to contrast our work with that in the existing literature, such as \citep{dino1,dino3}.

\subsection{Heating of the Earth's core}

\begin{figure}[htbp]
\centering
\resizebox{0.6\hsize}{!}{\includegraphics[scale=0.7]{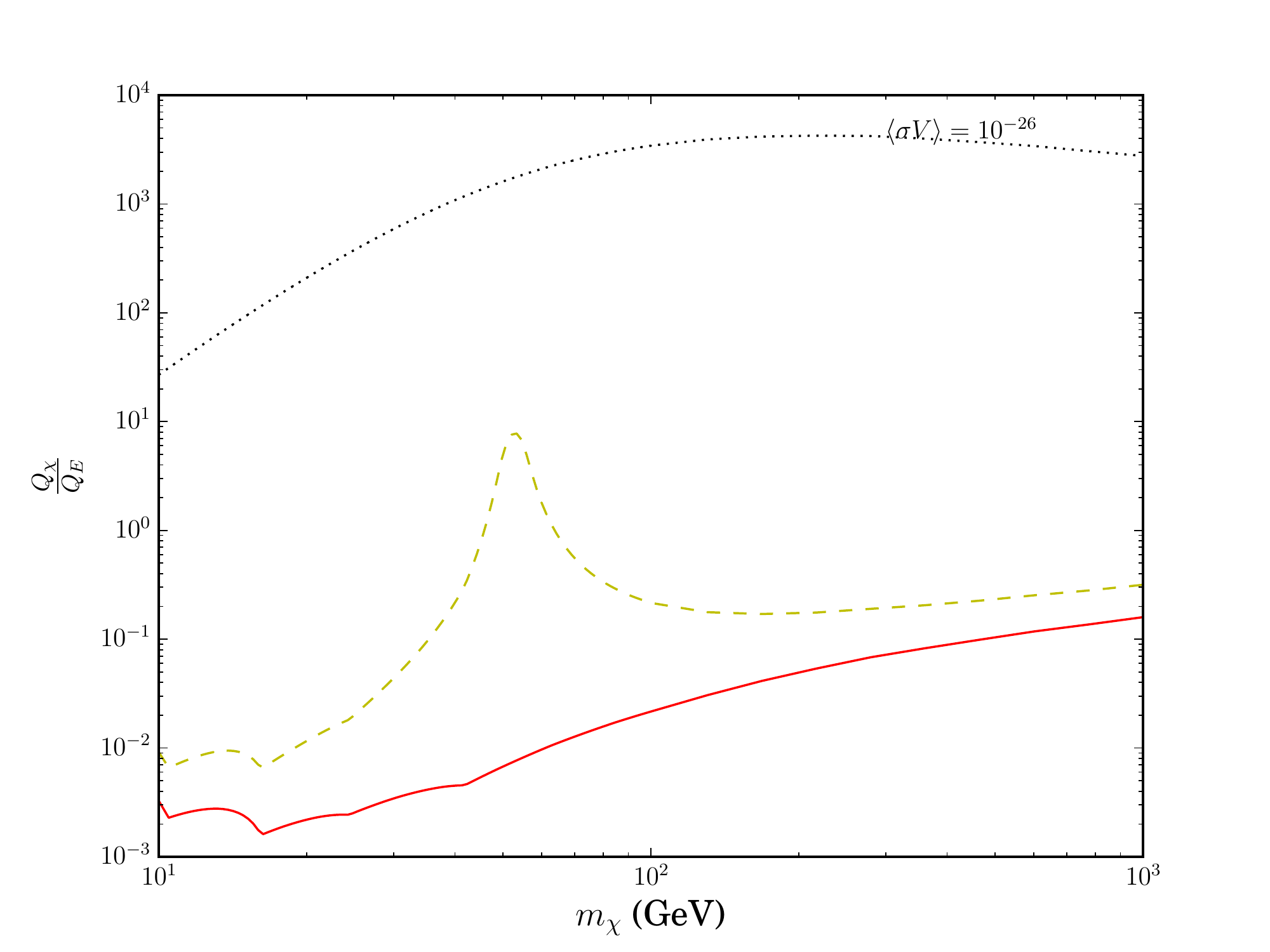}}
\caption{The ratio of the total heating produced by DM during UCMH transit to that produced normally within the core itself. This is independent of the mass of the UCMH chosen. The dotted black line shows the method of \citep{dino1,dino3}, the yellow dashed line shows the case using the capture rate from \citep{gould1987} with the LUX cross-sections, finally the solid red line shows the method of \citep{dino1} renormalised to the LUX experiment cross-sections. These were calculated assuming $\sigv = 10^{-26}$ cm$^3$ s$^{-1}$ and that Earth is composed only of iron. }
\label{fig:dQ}
\end{figure}

Figure~\ref{fig:dQ} shows the ratio of the total heat injected into the Earth's core via DM annihilation to that produced naturally by the mechanisms of radioactive decay, frictional heating, and residual heat from Earth's formation process. It is obvious that the use of the LUX experiment cross-sections limits drastically reduces the possibility of significant DM-induced heating. The method from \citep{dino1} with the more realistic UCMH model from \citep{bringmann2012}, summarised in Eqs.~(\ref{eq:rho}), (\ref{eq:size}), and (\ref{eq:rhomax2}), manages to produce significant heating with WIMPs at around $100$ GeV mass. The effects of the resonant scattering included in the method of \citep{gould1987} are also very strongly evident just below $100$ GeV mass WIMPs.

\begin{figure}[htbp]
\centering
\resizebox{0.6\hsize}{!}{\includegraphics[scale=0.7]{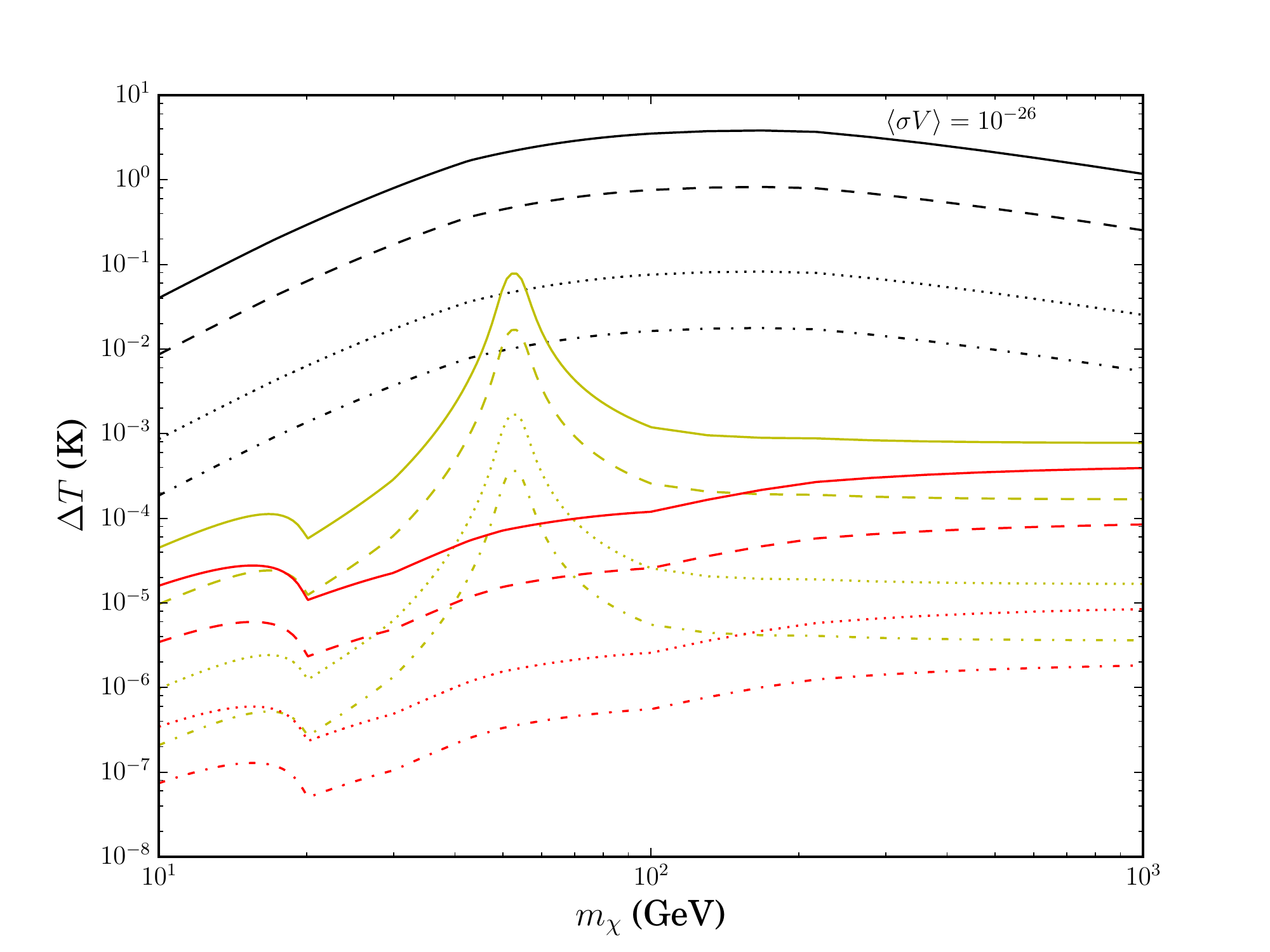}}
\caption{The total temperature change incurred within the Earth's core during UCMH transit. The black lines show the method of \citep{dino1,dino3}, the yellow show the case using the capture rate from \citep{gould1987} with the LUX cross-sections, finally the red lines show the method of \citep{dino1} renormalised to the LUX cross-sections. The  line style indicates the UCMH mass, solid lines are $10^3$ \msol, dashed are $10^1$ \msol, dotted are $10^{-2}$ \msol, and dot-dashed are $10^{-4}$ \msol. These were calculated assuming $\sigv = 10^{-26}$ cm$^3$ s$^{-1}$ and that Earth is composed only of iron.}
\label{fig:dT}
\end{figure}

Figure~\ref{fig:dT} shows the change in the temperature induced in the Earth's core during a UCMH transit for a wide range of UCMH masses. With the more detailed halo model used here we see that temperature changes on the order of $10^2$ K claimed in \citep{dino1,dino3} are unrealistic, even when we assume the same WIMP-iron scattering cross-section and large UCMH masses. The use of the LUX experiment cross-section limits rules out any temperature change above $10^{-2}$ K even for very large UCMH masses; we note that this maximum is achieved for the resonant scattering case. Without resonance, the maximal heating achieved is $10^{-4}$ K in the largest UCMHs considered. These temperature changes differ drastically from the order of magnitude figures given in \citep{dino1,dino3} and are  likely too small to even create significant hot-spots within the mantle layer of the Earth.
This result casts strong doubts upon the idea of DM-induced volcanism being responsible for mass extinction events, as the consequences of UCMH transit are not very dramatic. We find that the kind of UCMH needed to induce sufficient heating to achieve $10^2$ K changes will be above $10^5$ \msol. The abundance of such a large UCMH can be well constrained both by the analysis existing Fermi-LAT data~\citep{bringmann2012}, and in this work in Fig.~\ref{fig:fucmh_fermi_ska}. Since for WIMPs above 100 GeV, we can impose constraints on the abundance of $10^5$ M$_{\odot}$ UCMHs of $f < 3\times 10^{-4}$, we can determine that the most favourable periodicity for such a UCMH encounter by the solar system is $\tau \sim \mathcal{O}(10^2)$ Gyr. This suggests that there is at most a 1\% probability of a single such encounter within the lifetime of the universe, suggesting that even these large UCMHs could not account for periodic extinction events on Earth, but may be able to account for just one such episode. For smaller WIMP masses (below 100 GeV) this work cannot provide such limits from the current Fermi-LAT cross-sections, with the exception of annihilation via hard channels such as $\tau^+\tau^-$ (see Fig.~\ref{fig:fucmh_fermi_ska}), in which case we can extend the same limit on $f$ to these WIMP masses. This means only the soft annihilation channels like $b\bar{b}$ for WIMP masses below 100 GeV are not ruled out of explaining periodic mass extinctions with the current Fermi-LAT cross-sections by $\tau$ alone. The argument is, of course, stronger when our doubts about heat deposition magnitude are considered. Given that the $f$ constraints are only sensitive to cross-section limits below $10^{-29}$ cm$^3$ s$^{-1}$, with WIMP masses above 1 TeV (Fig.~\ref{fig:sigv_fmax}), it is unlikely that this exclusion can be weakened in the foreseeable future.

%

\begin{figure}[htbp]
\centering
\resizebox{0.6\hsize}{!}{\includegraphics[scale=0.7]{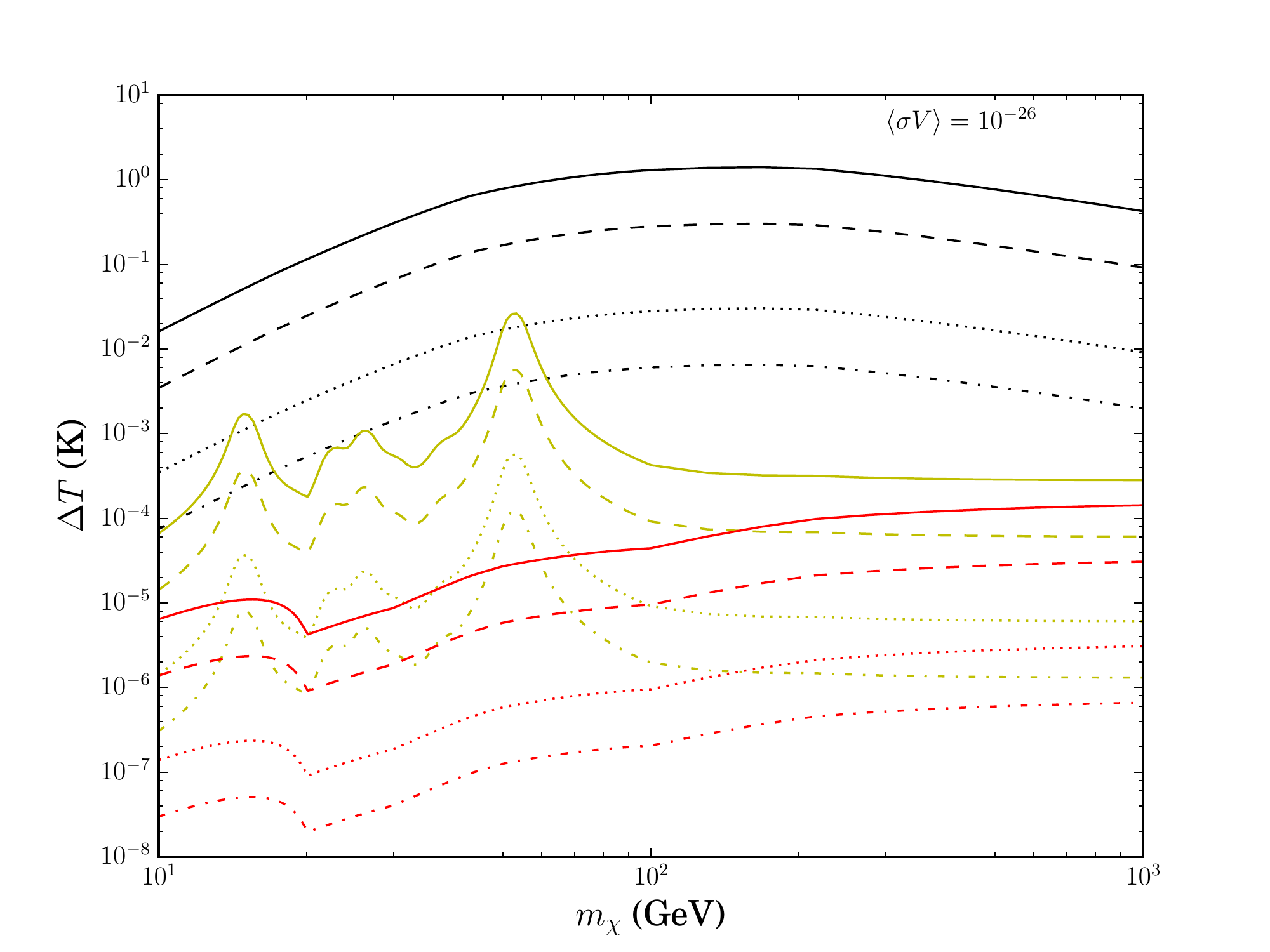}}
\caption{The total temperature change incurred within the Earth's core during UCMH transit with the advanced elemental composition model. The black lines show the method of \citep{dino1,dino3}, the yellow show the case using the capture rate from \citep{gould1987} with the LUX cross-sections, finally the red lines show the method of \citep{dino1} renormalised to the LUX cross-sections. The  line style indicates the UCMH mass, solid lines are $10^3$ \msol, dashed are $10^1$ \msol, dotted are $10^{-2}$ \msol, and dot-dashed are $10^{-4}$ \msol. These were calculated assuming $\sigv = 10^{-26}$ cm$^3$ s$^{-1}$.}
\label{fig:dTA}
\end{figure}

The same broad results are seen to hold true when we employ the advanced elemental composition model. The inclusion of these details even somewhat reduces the effect of DM annihilation heating, as shown in Fig.~\ref{fig:dTA}.

\subsection{Heating of the core of Mars}

\begin{figure}[htbp]
\centering
\resizebox{0.6\hsize}{!}{\includegraphics[scale=0.7]{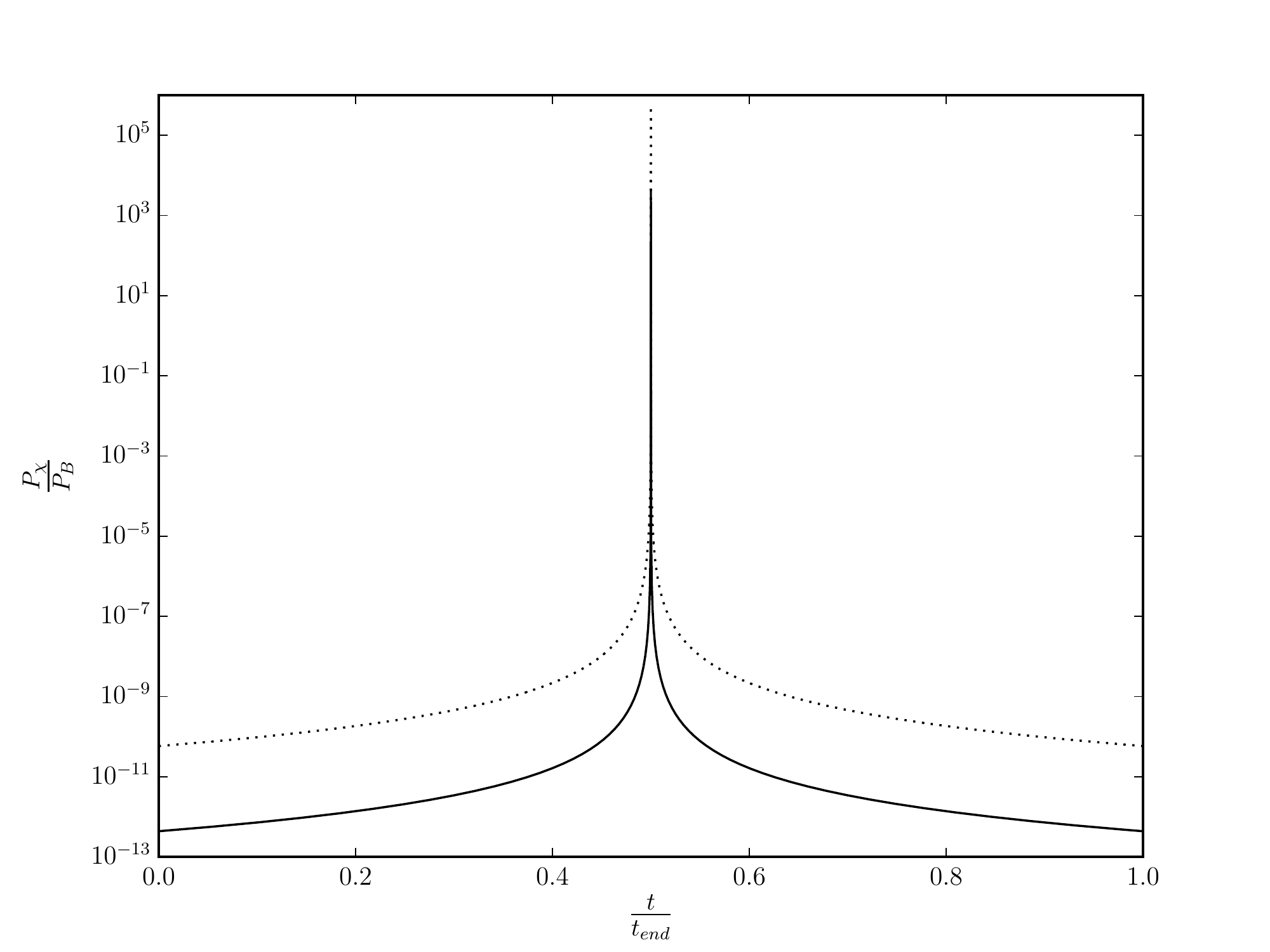}}
\caption{A comparison of the ratio of the power produced by DM annihilation during UCMH transit to the power emitted by Earth's core and the Martian geodynamo threshold. The solid line is for Earth and the dotted for Mars. This plot uses a halo mass of $10^3$ \msol, a WIMP mass of $1$ TeV, and an annihilation cross-section of $3 \times 10^{-26}$ cm$^3$ s$^{-1}$.}
\label{fig:power}
\end{figure}

Fig.~\ref{fig:power} shows the time dependence of the heating power of captured DM particles in terms of the ratio $\frac{P_{\chi}}{P_B}$, where $P_{\chi}$ is the DM-induced power output, and $P_B$ is the natural output of the given body capturing the DM particles. We use here the advanced elemental composition model. The DM heating power is normalised to the heat flow out of Earth's core ($P_B \approx 4 \times 10^{13}$ W~\citep{mack2007}) and the power-threshold for the geodynamo of Mars ($P_B \approx 3\times 10^{11}$ W~\citep{sandu2012}), respectively (this will correspond to the heat-flow within Mars core, as, in the de-gassing scenario, the heat-flow remains on the order of magnitude of the geodynamo threshold~\citep{sandu2012}). It is evident that the majority of the heating happens over a very short period of time, and that the effect is far more significant in the case of Mars than for the Earth.

\begin{figure}[htbp]
\centering
\resizebox{0.6\hsize}{!}{\includegraphics[scale=0.7]{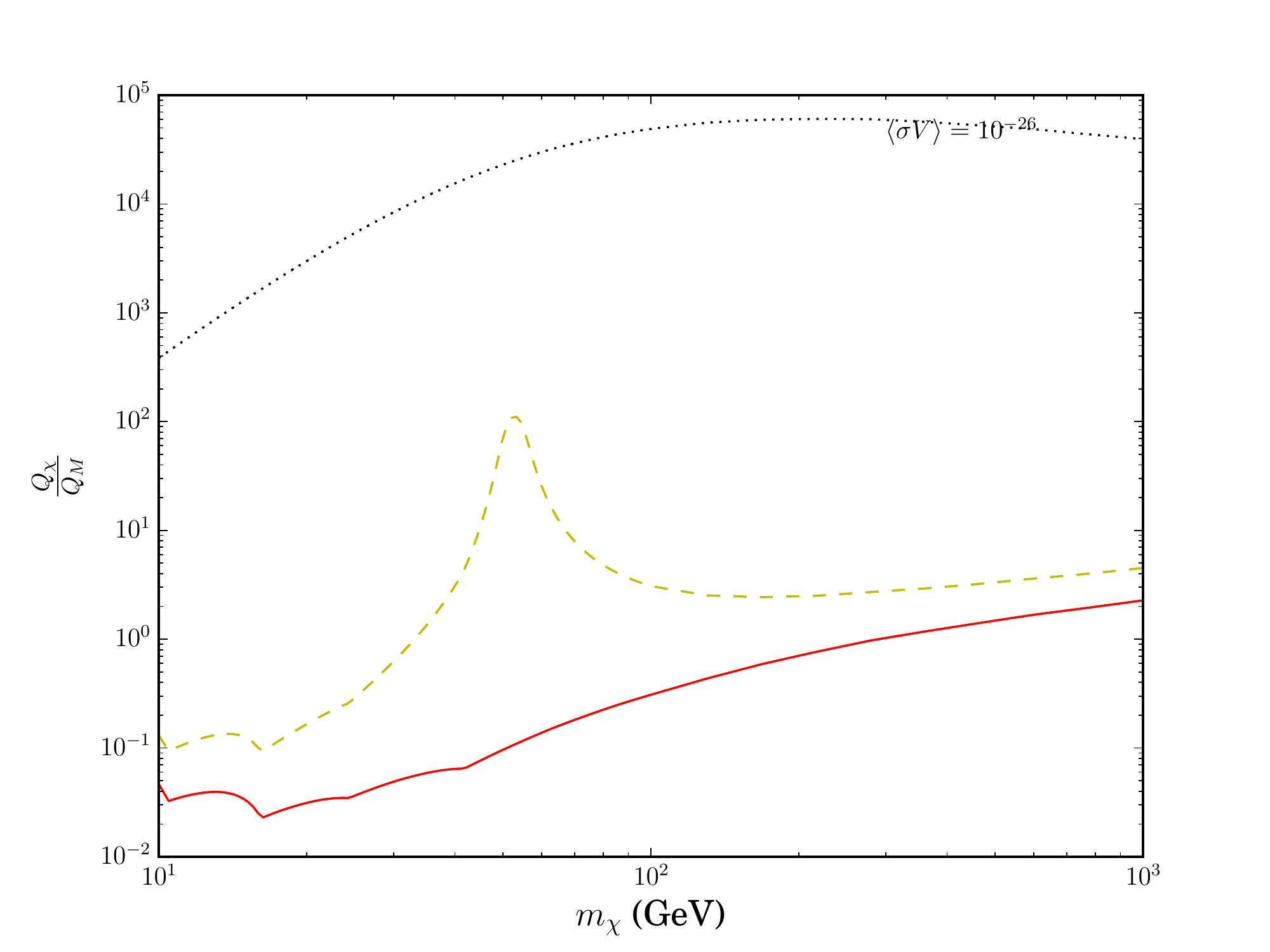}}
\caption{The ratio of the total heating produced by DM during UCMH transit to the threshold heat level for functioning of the geodynamo within Mars. This is independent of the mass of the UCMH chosen. The dotted black line shows the method of \citep{dino1,dino3}, the yellow dashed line shows the case using the capture rate from \citep{gould1987} with the LUX cross-sections, finally the solid red line shows the method of \citep{dino1} renormalised to the LUX cross-sections. These were calculated assuming $\sigv = 10^{-26}$ cm$^3$ s$^{-1}$ and that Mars is composed only of iron. }
\label{fig:dQ_mars}
\end{figure}

\begin{figure}[htbp]
\centering
\resizebox{0.6\hsize}{!}{\includegraphics[scale=0.7]{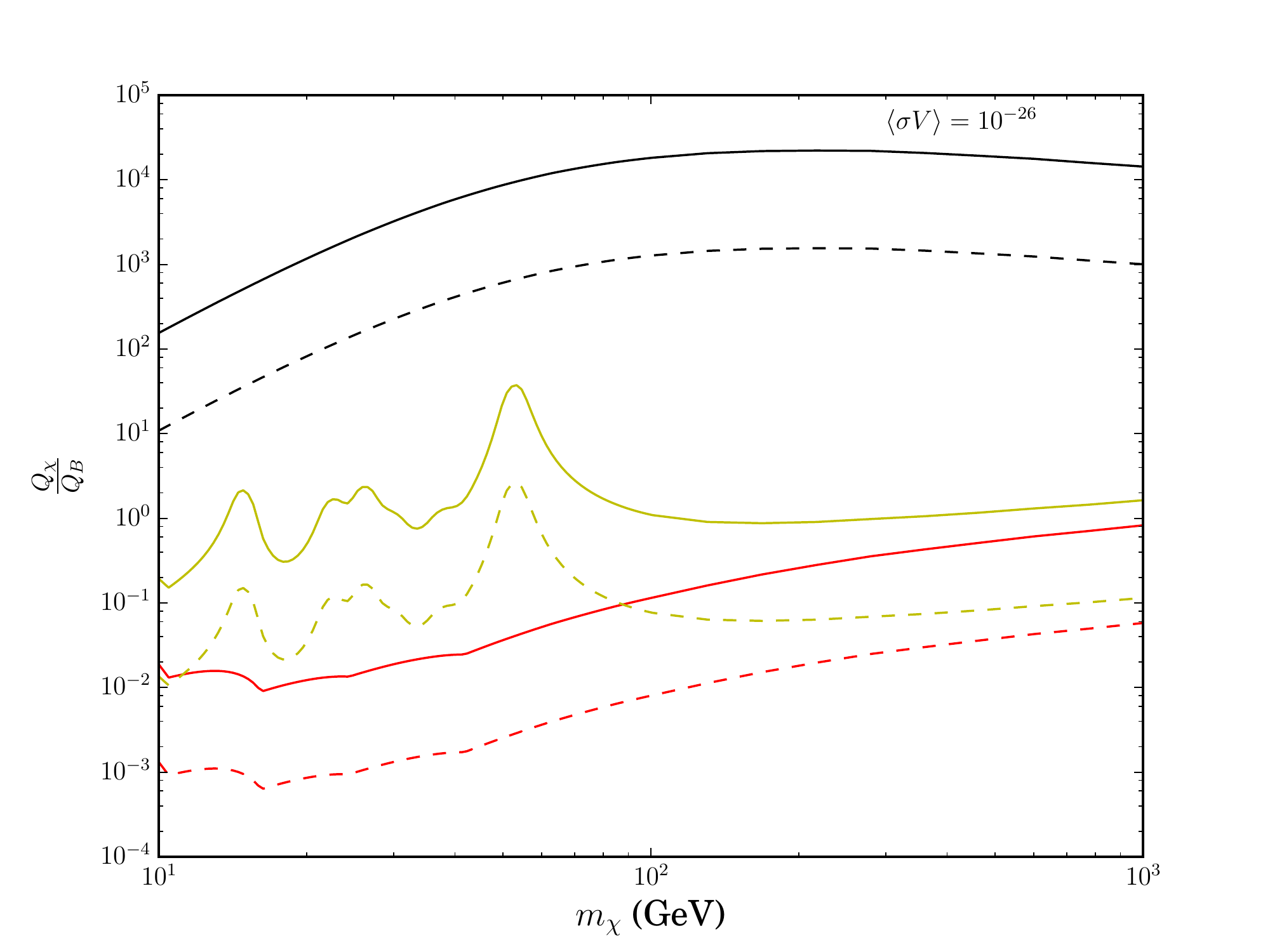}}
\caption{A comparison between Earth (dashed) and Mars (solid) of the ratio of the total heating produced by DM during UCMH transit to that of the heat emitted by Earth's core and the Martian geodynamo threshold, with the advanced elemental composition model. This is independent of the mass of the UCMH chosen.  The black line shows the method of \citep{dino1,dino3}, the yellow line shows the case using the capture rate from \citep{gould1987} with the LUX cross-sections, finally the red line shows the method of \citep{dino1} renormalised to the LUX cross-sections. These were calculated assuming $\sigv = 10^{-26}$ cm$^3$ s$^{-1}$. }
\label{fig:dQA_mars}
\end{figure}

Similar to Fig.~\ref{fig:dQ}, we display the total energy injected by DM annihilation into the core of Mars, normalised to the energy requirements of the geodynamo over the transit time (the power $P_B$ multiplied by $\Delta t$ the transit time), in Fig.~\ref{fig:dQ_mars}. Unlike the case of the Earth, the contribution of DM to the heat-flow is significant, even in the case of the LUX experiment scattering limits, being about one order of magnitude larger. This comparison is made directly for the advanced elemental composition case in Fig.~\ref{fig:dQA_mars}. This explicitly illustrates how much more significant the DM energy injection into the core is in the case of Mars.

\begin{figure}[htbp]
\centering
\resizebox{0.6\hsize}{!}{\includegraphics[scale=0.7]{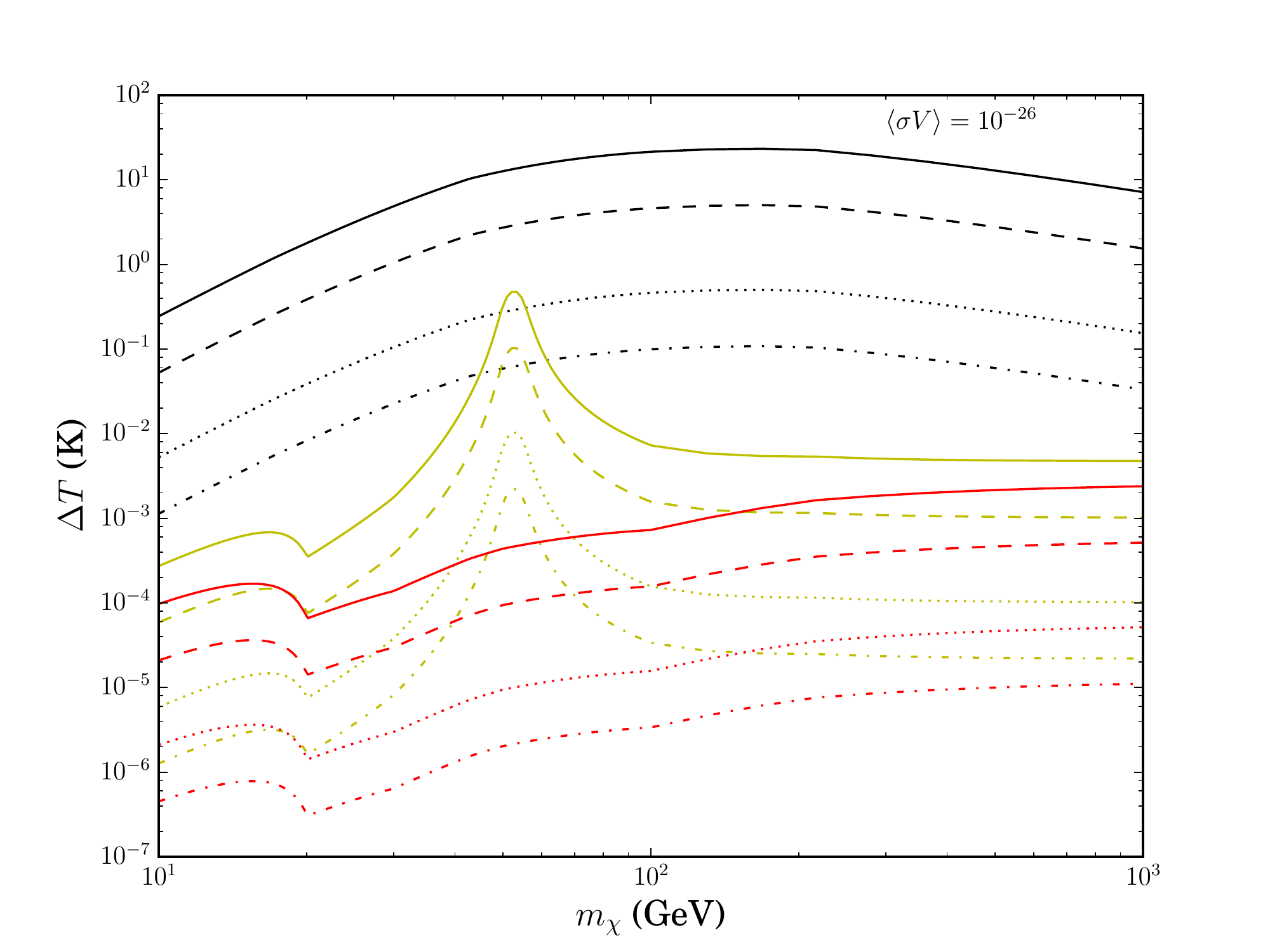}}
\caption{The total temperature change incurred within Mars' core during UCMH transit. The black lines show the method of \citep{dino1,dino3}, the yellow show the case using the capture rate from \citep{gould1987} with the LUX cross-sections, finally the red lines show the method of \citep{dino1} renormalised to the LUX cross-sections. The  line style indicates the UCMH mass, solid lines are $10^3$ \msol, dashed are $10^1$ \msol, dotted are $10^{-2}$ \msol, and dot-dashed are $10^{-4}$ \msol. These were calculated assuming $\sigv = 10^{-26}$ cm$^3$ s$^{-1}$ and that Earth is composed only of iron.}
\label{fig:dT_M}
\end{figure}

\begin{figure}[htbp]
\centering
\resizebox{0.6\hsize}{!}{\includegraphics[scale=0.7]{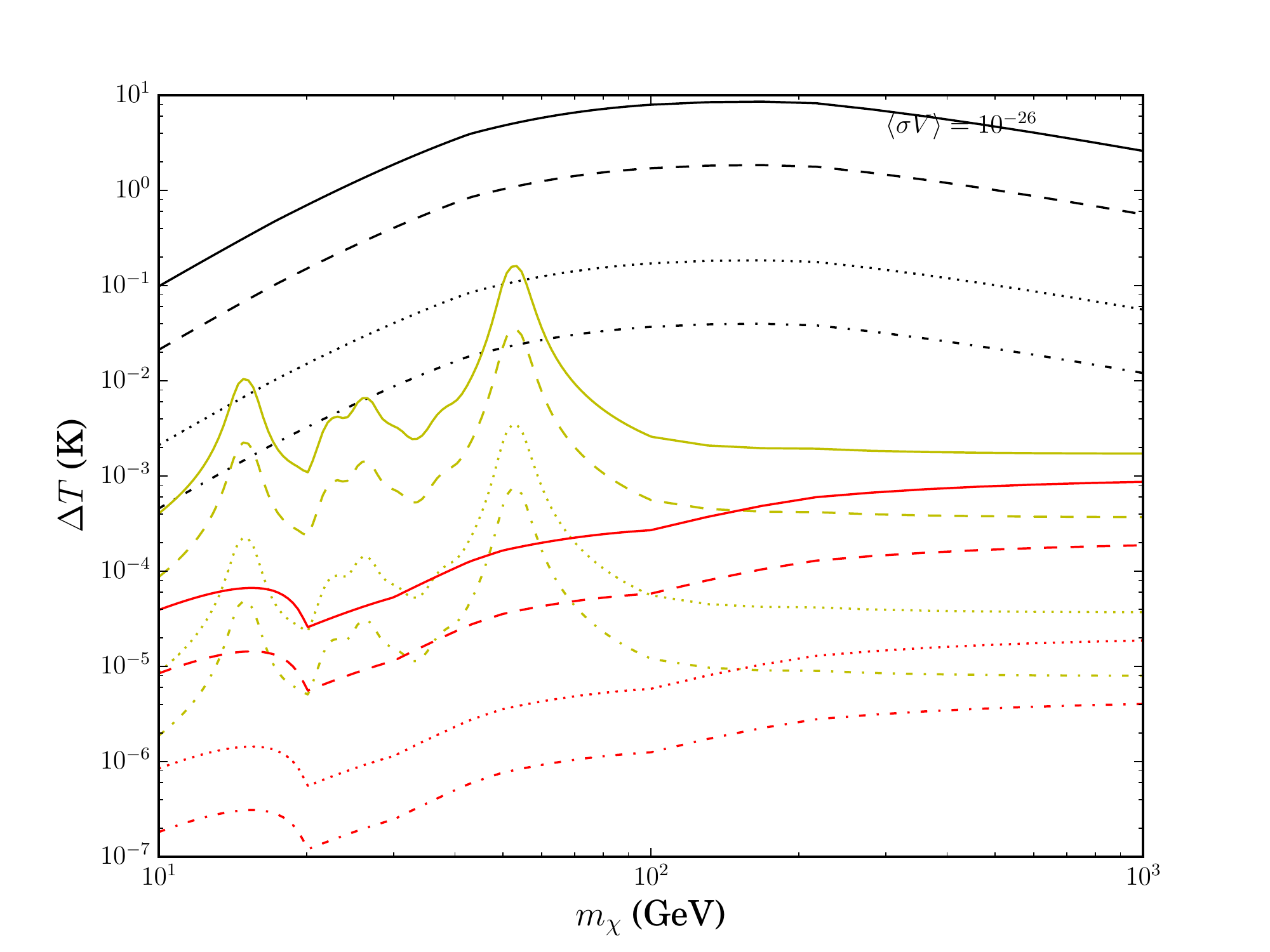}}
\caption{The total temperature change incurred within Mars' core during UCMH transit with the advanced elemental composition model. The black lines show the method of \citep{dino1,dino3}, the yellow show the case using the capture rate from \citep{gould1987} with the LUX cross-sections, finally the red lines show the method of \citep{dino1} renormalised to the LUX cross-sections. The  line style indicates the UCMH mass, solid lines are $10^3$ \msol, dashed are $10^1$ \msol, dotted are $10^{-2}$ \msol, and dot-dashed are $10^{-4}$ \msol. These were calculated assuming $\sigv = 10^{-26}$ cm$^3$ s$^{-1}$.}
\label{fig:dTA_M}
\end{figure}

Figures~\ref{fig:dT_M} and \ref{fig:dTA_M} show the temperature changes within the core of Mars for the iron-only and for the advanced composition models, respectively. In the case of the encounter geometrical cross-sections used in \citep{dino1,dino3} we see that a very significant heating of the order of 10 K is possible within the very short time, $\approx 0.1$ of the total transit time through the UCMH (which is $\mathcal{O} \lesssim 1$ year), that the bulk of the injection takes place. In the case of the LUX experiment limits, the change is reduced to $\sim 0.1$ K at most. This means that there is insufficient energy injected to create a significant change in the Mars core temperature, but that the level of heat flow and the temperature change are an order of magnitude more significant within Mars than within the Earth.

\subsection{Magnetic field loss on Mars}
\label{sec:mag}

Although an UCMH transit close to Mars does not deposit enough energy to change the temperature in Mars' core, there are other ways such an effect can cause the loss of life-sustaining conditions on the red planet. Mars is of particular interest here because it is thought to once have had an operating geodynamo which has since shut-off~\citep{jakosky2001,stevenson2001}. A possible explanation for the loss of geodynamo activity within Mars is provided in \citep{sandu2012}, where an argument is advanced that the Martian geodynamo can be shut down through a process of ``de-gassing" the planet's mantle. This effect is one of water and carbon-dioxide transfer from within the mantle to the planet surface, which can be extensive within Mars as the melt regions are far larger than those on Earth, and the red planet is lacking in active subduction processes that would re-gas the mantle, due to the stagnant Martian lithosphere. This effect can also be abetted by sudden volcanic activity within the mantle, as this can both de-gas through volcanic emission~\citep{grott2011}, but also, more generally, create new melt regions within the Martian mantle that transfer water to the lithosphere~\citep{sandu2012}. De-gassing can shut down the geodynamo as it changes the temperature required to maintain fluid flows within the mantle~\citep{sandu2012}, and combined with volcanic activity cooling the planet surface, this leads to the mantle solidifying and being unable to maintain a sufficiently rapid convection that is required to sustain the geodynamo.

Importantly, within the scenario of de-gassing, the heat flow convected out of the core of Mars is always on the order of magnitude of the minimum geodynamo heat-flow requirements~\citep{sandu2012}, as the rate of de-gassing rises with increasing heat-flow (as this stimulates more convection towards the lithosphere). This is significant as it is around $\sim 3\times 10^{11}$ W~\citep{nimmo2000,nimmo2007}, and is then two orders of magnitude below the convective heat-flow out of Earth's core. This means that Mars will be affected by a much more significant power injection, as illustrated in Fig.~\ref{fig:power}. The energy deposited by DM annihilation will be largely confined to the minimum of the gravitational potential and will be deposited very rapidly. This could serve to create an intense localised hot-spot on the core-mantle boundary. Such hot-spots would be very effective in expanding melt zones or creating rapid up-wellings of magma which provide a mechanism of increasing the de-gassing efficiency of the mantle. Such a process of hot-spot formation may also be possible on the Earth. However, without the possibility of strong de-gassing, and especially with the existence of re-gassing process at subduction zones, it is more difficult to link this to mass extinction events, although it could be associated with sudden, localized increases in volcanic activity on our planet.

Following the previous arguments, we propose a de-gassing process, boosted or initiated by DM, as a mechanism to explain the shut-down of the dynamo within the Mars' core. An absence of geodynamo activity would greatly reduce the Martial magnetic field and this then leads to the atmosphere of the planet becoming vulnerable to stripping by cosmic-ray winds and solar storms, as seen in \citep{maven}. Thus, although the volcanic activity stimulated by DM may not be enough to generate an extinction on its own, it may be sufficient to power the shut-down of the geodynamo, creating the conditions for an extinction event for any life-forms now exposed to dangerous radiation and absence of a protecting planetary atmosphere.
Additionally, this can also create a mechanism for the loss of water from the surface of Mars, as water vapour will be stripped along with the atmosphere, making an Earth-like water cycle unsustainable.\\
The period suggested in \citep{sandu2012} for the de-gassing  is around $1$ Gyr; if the periodicity for solar system encounters with UCMHs is taken to be $\approx 30$ Myr as suggested in \citep{dino3}, then this could result in as many as 30 UCMH transits for the the solar system within that de-gassing period, providing plenty of opportunity to aid in the de-gassing process. However, as discussed previously in Section~\ref{sec:capture}, the most optimistic periodicity derived in \citep{bringmann2012} from the Fermi-LAT galactic diffuse emission data is only on the order of $10$ Gyr for a 1 TeV WIMP with canonical cross-section. This seems to rule out an episodic contribution by UCMHs to the de-gassing scenario discussed here (at least for WIMPs above 100 GeV masses). However, there is still the opportunity for a UCMH encounter to trigger the de-gassing process initially by inducing a single temporary, but significant, increase to the heat output of the Martian core. Evidence for such a DM-induced de-gassing would have to be found within some powerful trigger event for Martian de-gassing process. This hypothesis can also be generally ruled out by stronger constraints on the abundance fraction $f$, for instance, if $f < 10^{-5}$ can be imposed on any UCMH mass between $10$ and $10^7$ M$_{\odot}$ for a whole range of WIMP masses, then the probability of the UCMH encounter, within the de-gassing time, is reduced to the level of $0.1$\% at most. In this work we have shown that the SKA and CTA could impose this constraint for all WIMPs above 100 GeVs for the halo masses $M_{UCMH} (0) < 3 \times 10^4$ M$_{\odot}$. Thus, at least one significant aspect of this hypothesis can be readily tested, over a broad parameter region, by future experiments.

\begin{figure}[htbp]
\centering
\resizebox{0.6\hsize}{!}{\includegraphics[scale=0.7]{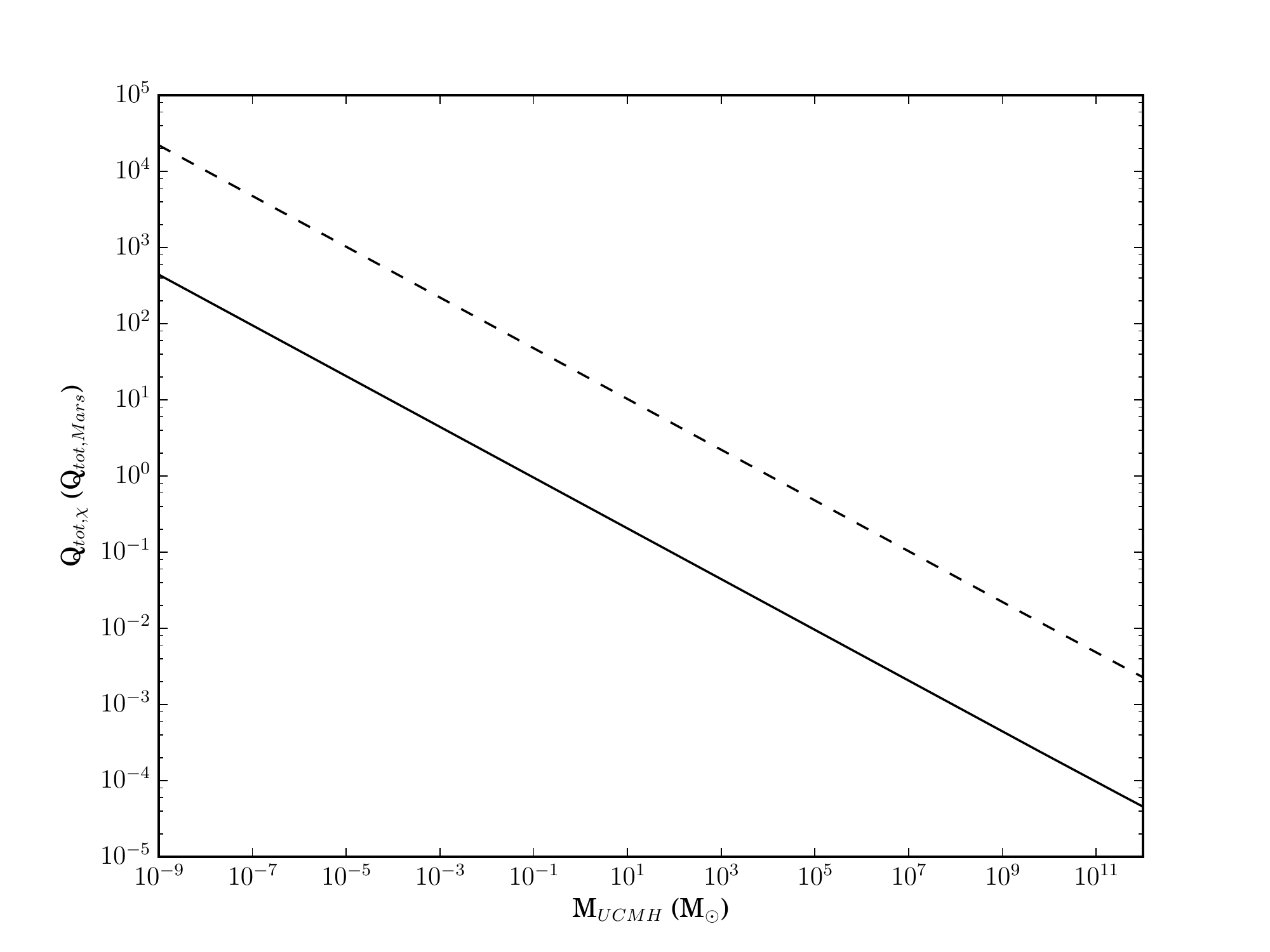}}
\resizebox{0.6\hsize}{!}{\includegraphics[scale=.7]{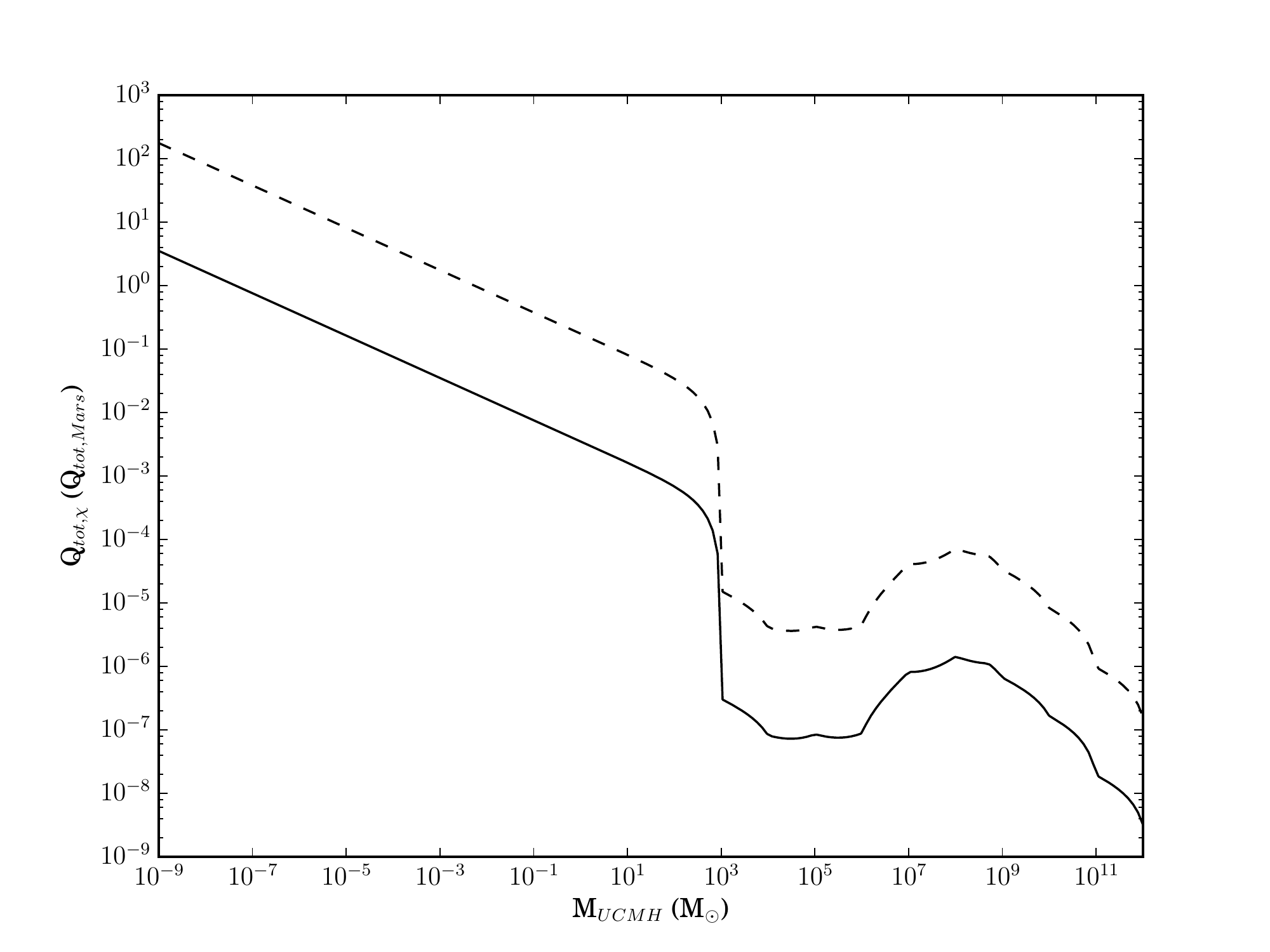}}
\caption{The energy deposited in the Martian core by DM capture (relative to that normally produced) during probable UCMH transits over a period of 1 Gyr. The solid line shows the LUX-normalised WIMP-nucleus scattering case, while the dashed line shows the normalisation used in \citep{dino1,dino3}. Top: we show a lensing-based upper-limit using $f \sim 0.1$ for a 1 TeV WIMP. Bottom: We use the Fermi-LAT limits on $f$ from \citep{bringmann2012} which apply to WIMPs of 1 TeV mass and canonical annihilation cross-section.}
\label{fig:tau}
\end{figure}

The issue of the encounter periodicity is summarised in Fig.~\ref{fig:tau}. Here we see that the probable total energy deposited by UCMH transits over 1 Gyr, relative to the normal energy output of the Martian core over this period. In the top panel we use a $1$ TeV WIMP with the micro-lensing upper-limit on compact DM of $f \sim 0.1$ and see that for a large range of halo masses we can manage a $\mathcal{O}(100)$ to $\mathcal{O}(10^{-1})$ contribution to core energy output, even with the WIMP-nucleon cross-section normalised to the LUX results. In the bottom panel we show the case of a 1 TeV WIMP with canonical cross-section using the existing Fermi-LAT $f$ limits from \citep{bringmann2012}. This clearly illustrates that the optimal halos for this de-gassing of Mars are small, having masses below $10^3$ M$_{\odot}$. The capture model from \citep{dino1,dino3} can result in nearly $\mathcal{O}(100)$ contributions to energy output of the Martian core for low mass halos. However, when we employ the LUX-normalised capture rate we see that the energy output fraction from DM is between $\mathcal{O}(1)$ and $\mathcal{O}(10^{-3})$ for masses below $10^3$ M$_{\odot}$. This means that the Martian UCMH de-gassing may remain viable for the 1 TeV WIMP with canonical cross-section. However, combined with the top panel, we can see that there is a substantial amount of room in the $f$ parameter space (around two to three orders of magnitude) that might allow for somewhat substantial contributions to the energy output of the Martian core. Additionally, it is notable that the Fermi-LAT constraints drop-off rapidly in severity for low mass halos~\citep{bringmann2012}, and in Fig.~\ref{fig:fucmh_fermi_ska} we see that, for Fermi-LAT dwarf search cross-sections, Fermi-LAT struggles to constrain the abundance of low mass halos, especially in the case of softer annihilation channels, like quarks, and for WIMP masses below 100 GeV. This means that there remain substantial regions of the parameter space needed for this de-gassing hypothesis that are not currently ruled out.

\section{Discussion \& Conclusions}
\label{sec:conclusions}

We have demonstrated that UCMHs will be both resolvable by experiments like SKA, as well as very strongly detectable. With larger mass UCMHs (corresponding to formation during the electron-positron annihilation epoch), within the radius set by accurate Gaia determination of the UCMH mass, being detectable above the milli-Jansky level with the SKA. This being a substantial improvement on the ability of Fermi-LAT to locate these objects. Additionally, this implies that objects like UCMHs could provide a window on remote cosmological epochs via the SKA.

Additionally, we have shown that the dual search methodology we have proposed allows for a robust link between experimental limits on DM annihilation cross-sections and the abundance of UCMHs. This was done by proving a lack of significant degeneracy between the abundance and cross-section limits within a region limited by requirement of correlating potential indirect DM emissions with unknown dark objects observed via microlensing. At the same time we demonstrated that within this search region we can expect experiments like SKA and CTA to be able to stringently probe UCMH abundances even for very small values of the WIMP annihilation cross-section ($\sim 10^{-29}$ cm$^3$ s$^{-1}$). Prospects for decaying DM detection within UCMHs were, in contrast, extremely weak.

We further explored the effects of imposing the $\langle \sigma V \rangle$ limits from Fermi-LAT observations of dwarf spheroidal galaxies combined with a the lack of reported UCMH candidate observations (which also means a lack of follow-up targets for Gaia in the search region). We found that these limits are around $f_{max} \sim 10^{-7}$ for halos with mass $10^3$ \msol across the WIMP mass range, $10$ GeV to $10$ TeV, in agreement with the results of \citep{bringmann2012} at 1 TeV with cross-sections close to the canonical relic value. Despite the high sensitivity of the SKA, a lack of UCMH candidate detections place similar limits on the abundance. However, these can extend, with a similar stringency, down to smaller halo masses and over a larger variety of annihilation channels than for Fermi-LAT. In the case of annihilation channels with softer emission spectra (such as $b\bar{b}$), the SKA cannot constrain UCMH abundances when WIMP masses are below 100 GeV. For decaying DM, Fermi-LAT cannot set abundance limits through the non-observation of UCMHs when the decay-rate limits from Fermi-LAT extra-galactic data are imposed. For the SKA, limits cannot be set on any halos of $\le 10^3$ $M_\odot$ for the same decay rate limits.

The final part of this analysis was a derivation of the limits on $f_{max}$ using the value of the null-constraints on $\sigv$ that would arise from non-observation of radio and $\gamma$-ray signals in the Reticulum II dwarf galaxy. These were then used to impose constraints on $f_{max}$ assuming that each of the CTA, SKA, and Fermi-LAT telescopes do not observe UCMH candidates within the region Gaia can accurately characterise targets. The SKA was limited for WIMPs below 100 GeV when softer annihilation channels are considered, but was able to extend this analysis down to smaller halo masses than Fermi-LAT, especially in the case of harder annihilation channels like $\tau$ leptons. CTA displayed similar results to Fermi-LAT.
Radio constraints from a source like Reticulum II are going to be affected by confusion limits and discrete source subtractions. Far simpler constraints are derivable from non-observation of $\gamma$-rays in Reticulum II, as dwarf galaxies lack processes likely to produce significant $\gamma$-ray emissions. However, poor spatial resolution that may lead to source confusion suggests a combined strategy of supplying cross-section limits from very high energy observation and hunting faint UCMHs through correlation of the highly sensitive SKA and microlensing probes like Gaia.

In the local environment of our solar system, the hypothesis of volcanogenic DM, connecting UCMHs to mass extinctions of life on Earth was explored. We have shown that the heat-flow resulting from DM capture within Earth's core is far less significant than has been argued with less rigorous models of UCMH formation and density~\citep{dino1,dino3}. Additionally, the hypothesis is greatly weakened when the nuclear scattering cross-section is renormalised to reflect the bounds on WIMP-nucleon interactions from the LUX experiment, and the elemental composition of the Earth is accounted for. Additionally, existing constraints on the UCMH abundance $f$ imply that the periodicity of UCMH encounters with the solar system is at best $\tau > \mathcal{O}(10)$ Gyr for UCMHs above $10^2$ M$_{\odot}$ (with a 1 TeV mass WIMP in \citep{bringmann2012} that we extended down 10 GeV), far larger than the 30 Myr value which is suggested to fit these occurrences on Earth~\citep{dino2}. However, the far lower heat-flow required within the core of Mars makes it a far stronger candidate for linking DM to mass extinction.
We have shown that the heating from DM capture and annihilation is orders of magnitude more significant within Mars than Earth, and its very rapid deposition implies that high-temperature hot spots can be induced on the core-mantle boundary. This is linked to the extinction of Martian life through the hypothesis that rapid volcanic and melt-zone activity within Mars can contribute to a process of ``de-gassing" the mantle~\citep{sandu2012}. This process is one of removing water and other light gases from the mantle, either through volcanism, or loss into the lithosphere via magmatic melt-zones. The end result is that a larger temperature is needed to maintain a fluid mantle state, thus, leading to feedback that induces mantle solidification and a shut-down of the convection cycle needed to sustain geodynamo activity.
The final outcome of this process is a systematic decrease of the magnetic field in Mars that allows cosmic rays and the solar wind to strip Mars of its atmosphere and surface water, making a life-cycle unsustainable. This is a possibility due to the nature of the boundary between the mantle and lithosphere within Mars, which has extended melt-zones under the stagnant crust, and, unlike Earth, does not possess a significant lithosphere recycling process that would re-gas the mantle~\citep{sandu2012}. The fact that the time-scale of de-gassing is around 1 Gyr means that there is little chance of repeated UCMH encounters with the solar system during this time except for low-mass halos. This means that low-mass, $M_{UCMH} < 10^2$ M$_{\odot}$, UCMHs cannot be ruled out as powering a Martian de-gassing process based on present UCMH abundance limits (for WIMPs below 100 GeV in mass).
Evidence for this hypothesis would have to be found in sudden or episodic increases in the activity of the Martian mantle during the period of the geodynamo loss. One method of excluding this hypothesis would also be to impose stringent constraints on the UCMH abundance $f$. If $f < 10^{-5}$ can be imposed on any UCMH mass between $10$ and $10^7$ M$_{\odot}$ for a whole range of WIMP masses below 100 GeV, then the probability of the UCMH encounter is reduced to the level of $0.1$\% at most. In this work we have shown that the SKA and CTA could impose this constraint for all WIMPs above 100 GeVs for the halo masses $M_{UCMH} (0) < 3 \times 10^4$ M$_{\odot}$. For hard annihilation channels, like $\tau$ leptons, this future constraint on $f$ can be extended down to $10$ GeV WIMPs. This leaves un-excluded the possibilities of smaller UCMHs, or $m_{\chi} < 100$ GeV models annihilating via soft annihilation channels, like via quarks, playing a role in the de-gassing history of Mars.
%

Given the theoretical motivation for existence and persistence of UCMH objects into the present epoch, the experimental hints at microlensing by compact objects, and the wide-ranging consequences of their existence and abundance, it is important to be able to make effective searches for these objects. The results presented here demonstrate that searches linked to other indirect DM detection experiments are able to produce robust and meaningful results with significant local and cosmological consequences. This is provided the methodology of requiring correlated microlensing and indirect detections of unknown dark objects is imposed to provide a physically motivated limited search radius. With this search technique in place, the constraints on the $\sigma V - M_{\chi}$ plane derived from future studies of dwarf galaxies in radio and $\gamma$-rays can then be used to impose constraints on the fraction of DM found in UCMHs, and these can be especially stringent and applied to wide range of UCMH masses if no candidates are located by joint observations with experiments such as the SKA and Gaia. However, only for large halo masses $> 10^3$ M$_{\odot}$ can we similarly constrain the abundance $f$ in the case of exclusively decaying DM, this comes with the additional restriction that decaying DM abundance limits cover one decade of WIMP mass fewer than those of annihilation.

In summary, we have demonstrated a wide-ranging effects of UCMHs, from constraining over-all cosmology as well as structure formation within our galaxy, to affecting the geological history of solar-system planets. We have also shown that robust indirect detection constraints on the abundance of these objects is indeed possible, provided we correlate this search with microlensing data. The study of these peculiar halos with the next generation astronomical telescopes will offer a rich data set to constrain their properties. This places these micro-structures on multiple frontiers of modern cosmology.

\section*{Acknowledgements}
SC acknowledges support by the South African Research Chairs Initiative
of the Department of Science and Technology and National
Research Foundation, as well as the Square Kilometre Array (SKA).
This work is based on the research supported by the South African
Research Chairs Initiative of the Department of Science and Technology
and National Research Foundation of South Africa (Grant
No 77948). GB acknowledges support from the DST/NRF
SKA post-graduate bursary initiative.

\section*{References}
\bibliographystyle{elsarticle-harv}
\bibliography{ucmh}

\end{document}